\documentclass[a4paper,11pt]{article}
\usepackage{jheppub}
\usepackage{hyperref}

\newcommand{\be}{\begin{equation}}
\newcommand{\ee}{\end{equation}}
\newcommand{\pa}{\partial}
\newcommand{\calR}{{\cal R}}
\newcommand{\vphi}{\varphi}
\newcommand{\deltaV}{\delta V}
\newcommand{\dprime}{{\prime\prime}}
\newcommand{\rs}{u} %
\newcommand{\morder}[1]{\mathcal{O}\left(#1\right)}

\title{%
Large-$D$ approach to %
nearly critical holographic matter and Yang--Mills deconfinement
}

\author[a]{Matti Järvinen,}

\author[b,c]{Leonardus B.~Putra,}

\author[d]{and Dorin Weissman}

\affiliation[a]{Institute of Theoretical Physics, Chinese Academy of Sciences, Beijing 100190, China}

\affiliation[b]{Mathematical Institute, University of Oxford, Radcliffe Observatory, Andrew Wiles Building, Woodstock Rd, Oxford OX2 6GG}

\affiliation[c]{Centre for Quantum Fields and Gravity, Department of Physics, Swansea University, Swansea SA2 8PP,
United Kingdom}

\affiliation[d]{INFN Sezione di Napoli, Monte S. Angelo, Via Cintia, 80126 Naples, Italy}

\emailAdd{mattijarvinen@itp.ac.cn}
\emailAdd{2635236@swansea.ac.uk}
\emailAdd{dorin.weissman@na.infn.it}

\abstract{
We study a systematic expansion of dynamical black hole solutions in a class of five-dimensional Einstein--dilaton theories that are holographically dual to nearly critical, strongly coupled four-dimensional field theories. The expansion is motivated, through dimensional reduction, by the large-$D$ limit of pure gravity, where $D$ is the number of spacetime dimensions. At leading nontrivial order, the Einstein equations reduce to a closed set of non-relativistic hydrodynamic equations in four dimensions, as first shown in our previous letter; these equations generalize the large-$D$ effective theory of black branes. We derive the thermodynamics and hydrodynamics of the system from black hole thermodynamics and the holographic dictionary, and show that the energy-momentum tensor extracted from the near-boundary geometry is conserved precisely when the effective equations hold. We analyze planar, cylindrical, and spherical domain walls between coexisting phases, as well as non-equilibrium steady-state walls with nonzero momentum flux. We also show how the near-horizon geometry can be completed by a holographic flow to an asymptotically AdS$_5$ region. Finally, as an application to the confinement–deconfinement transition of Yang--Mills theory, we construct the domain wall between the confined and deconfined phases in improved holographic QCD, compute its surface tension, and compare it with recent lattice results.
}

\usepackage{xcolor}

\begin{document}
\maketitle

\flushbottom

\clearpage
\section{Introduction}
Gauge-gravity duality provides a powerful framework for computing observables of strongly coupled gauge theories by studying classical gravitational theories in a curved spacetime with one dimension more than the dual field theory. Nevertheless, determining the gravitational dynamics remains a challenging task. Except in highly symmetric or static settings, one typically has to resort to computationally intensive numerical methods to solve the large system of coupled nonlinear partial differential equations arising from the Einstein equations.

Developing controlled approximation schemes that retain the essential physics while reducing the computational cost is therefore of considerable interest, especially when studying solutions with time-dependent dynamics or with nontrivial spatial structure, such as domain walls and bubbles. A remarkable simplification occurs in the large-$D$ limit of general relativity \cite{Emparan:2013moa,Emparan:2020inr}, where $D$ denotes the spacetime dimension. In this limit, the Einstein equations describing dynamical black hole (or more generally, black brane) geometries reduce to a much smaller set of equations governing a few effective degrees of freedom \cite{Emparan:2015gva,Emparan:2015hwa,Emparan:2025yfy}. These equations define a non-relativistic hydrodynamic effective theory, in which the dynamics of the black hole are captured by a thin membrane localized near the black hole horizon~\cite{Bhattacharyya:2015dva,Dandekar:2016jrp}.

Moreover, the effective equations take a form with only linear time derivatives, which makes simulations much simpler and more straightforward than simulating Einstein gravity~\cite{Emparan:2015gva,Andrade:2018yqu,Emparan:2021ewh,Licht:2022rke,Luna:2022tgh,Emparan:2023dxm}. At large~$D$, the hydrodynamic sector is decoupled from other modes~\cite{Emparan:2014aba}, such that expanding in $1/D$ automatically leads to a hydrodynamic theory. This is different from the fluid-gravity correspondence~\cite{Hubeny:2011hd}, where one introduces an explicit derivative expansion.

At first glance, the large-$D$ limit may appear to have little to do with holographic descriptions of lower-dimensional gauge theories, since holography typically relates gravity in $D$ dimensions to a field theory in $D-1$. Remarkably, however, the large-$D$ limit is particularly well suited to the study of the confinement-deconfinement transition in four-dimensional Yang--Mills theory. Upon dimensional reduction to five dimensions and after consistently decoupling the gravitational fluctuations associated with the transverse directions, Einstein gravity at large $D$ gives rise to an Einstein-dilaton theory with a specific dilaton potential~\cite{Kanitscheider:2008kd,Gouteraux:2011qh}. Within the framework of Improved Holographic QCD (IHQCD)~\cite{Gursoy:2007cb,Gursoy:2007er}, this potential is precisely the critical potential separating confining and deconfining phases. Specifically, after the reduction of Einstein gravity on pure $AdS_{D}$ to five dimensions, the effective theory is an Einstein-dilaton theory with the potential $V(\phi) \to V_0 e^{\frac43(\phi-\phi_0)}$ in the $D\to\infty$ limit. The exponent $\frac43$ is precisely the critical value where the resulting geometry starts corresponding to a confining dual field theory \cite{Gursoy:2007er}, and hence there is a highly non-trivial correspondence of the large-$D$ limit and the confinement-deconfinement transition (see~\cite{Betzios:2018kwn,Gursoy:2021vpu}). In fact, this observation holds more generally also for spin models near criticality~\cite{Gursoy:2010kw} and a wider class of theories which have a higher order phase transition at finite temperature~\cite{Gursoy:2010jh}.

The novelty of our present approach, introduced in  \cite{Jarvinen:2024vdi}, is that, rather than restricting ourselves to the specific Einstein-dilaton theory obtained by dimensional reduction, we consider arbitrary perturbations of the dilaton potential about its infinite $D$ limit. In this way, our model is formulated directly in five dimensions, with $1/D$ serving as a controlled perturbative expansion parameter. We construct the effective theory to leading order in $1/n$ corrections, where $D = 5+n$, so that $n$ denotes the number of dimensions that are integrated over in the dimensional reduction. We show that we can write a simple generalization of the large-$D$ effective theory of \cite{Emparan:2015gva} and related works, when the $\frac1n$ correction of the potential is given by an arbitrary function $\deltaV$.

In addition, we derive simple expressions for the black hole thermodynamics and construct the energy-momentum tensor for general configurations with corrections from an arbitrary $\deltaV$. The conservation equations of the energy-momentum tensor reproduce the effective theory. The entropy density and temperature determine the speed of sound and the free energy density through standard thermodynamic relations. This approach provides a consistent thermodynamic description of the effective theory.

Domain walls and bubbles are central to the real-time dynamics of first-order phase transitions, which have been studied extensively in holography by solving the full Einstein equations numerically~\cite{Janik:2015iry,Attems:2017ezz,Attems:2019yqn,Bellantuono:2019wbn,Bantilan:2020pay,Attems:2018gou,Bea:2021zsu,Bea:2022mfb,Hoyos:2026eoq}, and by effective approaches~\cite{Janik:2021jbq,Janik:2022wsx,Ares:2021ntv,Ares:2021nap,Henriksson:2025vci,Janik:2025uqa}.
We are particularly interested in domain wall solutions that interpolate between phases~\cite{Aharony:2005bm,Figueras:2014lka} (see also~\cite{Berenguer:2026ftk}). We write the general time-dependent equations that can describe the evolution of planar domain walls and cylindrical and spherical bubbles, relevant for first-order phase transitions in the early universe and the resulting gravitational waves~\cite{LISA:2022yao,NANOGrav:2023gor}, which have also been studied holographically~\cite{Bigazzi:2020avc,Ares:2020lbt,Bigazzi:2021ucw,Bea:2021zsu,Bea:2021zol}. For static solutions, we derive conditions with simple thermodynamic interpretations, including the effect of surface tension for curved interfaces. We also discuss stationary solutions with nonzero momentum flow, which can be interpreted as duals of non-equilibrium steady states (NESS)~\cite{Bhaseen:2013ypa,Pourhasan:2015bsa,Herzog:2016hob,Sonner:2017jcf,Ecker:2021ukv,Bachas:2021tnp}.

As a concrete application of the model, we study domain wall solutions in the context of improved holographic QCD~\cite{Gursoy:2007cb,Gursoy:2007er,Gursoy:2008za,Gursoy:2009jd,Gursoy:2010fj} and compare our results with lattice data for Yang–Mills theory~\cite{Panero:2009tv}. Comparisons with lattice thermodynamics and the full holographic model provide indirect checks of the accuracy of the large-$D$ approximation, while the calculations themselves are numerically inexpensive compared with lattice simulations or solutions of the full holographic model. Although this application does not involve an explicit large-$D$ parameter, the potential has a numerically large overall coefficient $V_0$ and small relative corrections in the relevant regime, motivating the use of the expansion. Within this framework, we construct domain walls for several choices of the dilaton potential and determine their widths and surface tensions, exploring their dependence on the potential and comparing the surface tensions with lattice results~\cite{Lucini:2005vg,Rindlisbacher:2025dqw}.

The outline of this paper is as follows. In section~\ref{sec:model} we define our model and write the Ansatz for dynamical black hole solutions. In section~\ref{sec:BHET} we present the effective theory, including the derivation of the equations and the black hole thermodynamics. Section~\ref{sec:dw} describes domain wall solutions, cylindrical and spherical bubbles, and the NESS solutions. In section~\ref{sec:UV} we discuss the UV completion of the effective theory and the matching of the near-horizon dynamics to fluctuations in the far region (see~\cite{Emparan:2014aba,Suzuki:2015axa,Emparan:2018bmi} in the large-$D$ literature). Finally, in section~\ref{sec:application}, we apply our approach to Yang--Mills theory, including comparisons of our predictions to lattice results.

The paper also includes several technical appendices. In Appendix~\ref{app:dimred} we briefly review the dimensional reduction. In appendices~\ref{app:nlosolution} and~\ref{app:UVasymptotics}, respectively, we write the explicit next-to-leading order solution of the background geometry and its near boundary asymptotic form. Appendix~\ref{app:FG} provides the full technical details on the derivation of the energy-momentum tensor.

\section{The large D inspired framework} \label{sec:model}

We consider five-dimensional Einstein-dilaton gravity,
\begin{equation} S = \frac{1}{16\pi G_5}\int d^5x \sqrt{-g}\left(R - \frac43(\pa \phi)^2 + V(\phi)\right) \,,\label{eq:EDaction} \end{equation}
with a specific choice of the potential
\begin{equation} 
 V(\phi) = V_0 e^{\frac43\left(\phi-\phi_0\right)}\left(1+\frac1n \delta V(\phi-\phi_0) + {\mathcal O}\left(\frac1{n^2}\right)\right) \,.\label{eq:potential}
\end{equation}
The potential depends on three parameters $V_0$, $\phi_0$, and $n$, and we are considering the limit where $n \to \infty$.
The function $\delta V(\phi)$, which parametrizes a small perturbation to the potential at large $n$, is left as an arbitrary function of $\phi$.

As explained in~\cite{Jarvinen:2024vdi}, the Ansatz~\eqref{eq:potential} is motivated by a dimensional reduction from $D=5+n$ dimensional gravity. As we show in Appendix~\ref{app:dimred}, for specific choices of $V_0$ and linear $\deltaV$, our setup is a consistent reduction of specific large-$D$ setups. However, we choose to work in five dimensions and consider an arbitrary perturbation $\deltaV$ instead of the choices arising from dimensional reduction.

We wish to consider dynamical black hole solutions in our framework. To do this, following the large-$D$ approach~\cite{Emparan:2015gva,Emparan:2016sjk}, we first write an Ansatz for the metric in Eddington--Finkelstein coordinates, boosted along the spatial directions $x^i$:
\begin{equation} ds^2 = e^{2A}\left(-f dt^2 - 2\,dt\,dr - 2 C_i\, dt\, dx^i + g_{ij} dx^i dx^j\right) \label{eq:gds2}\,. \end{equation}
This form also consistently fixes the gauge; see~\cite{Andrade:2018zeb} for a discussion on this.
Including the leading order and next-to-leading order terms in $1/n$, the Ansatz can be written as
\begin{align} \label{eq:AAnsatz}
A({\cal R},t,x^k) &= A_0-\frac{n r}{3\ell} + \frac 1n A^{(1)}({\cal R},t,x^k)\,, \\
f({\cal R},t,x^k) &= 1 - m(t,x^k) \calR + \frac{1}{n} f^{(1)}({\cal R},t,x^k)\,,
\label{eq:fAnsatz}
\\
g_{ij}({\cal R},t,x^k) &= \frac{1}{n}\delta_{ij} +\frac{1}{n^2}\frac{p_i(t,x^k) p_j(t,x^k)}{m}{\cal R}\,, \label{eq:gijAnsatz}
\\
C_i({\cal R},t,x^k) &= \frac{1}{n} p_i(t,x^k) \calR 
+ \frac{1}{n^2}C_i^{(1)}(\calR,t,x^k)\,, \label{eq:gci}
\end{align}
and is defined in terms of
\begin{equation} \label{eq:calRdef}
\calR \equiv e^{\frac{n r}{\ell}} \,,
\end{equation}
which should be kept fixed as $n \to \infty$ in order to capture the dynamics of the black holes.
The holographic boundary is therefore located at $r \to -\infty$ where ${\cal R} \to 0$. The black hole horizon is located at ${\cal R} = 1/m(t,x^k) + \mathcal{O}\left({1/n}\right)$.

The constant $V_0$ is parameterized as
\begin{equation} V_0 = \frac{n^2}{\ell^2}e^{-2A_0}\,. 
\label{eq:V0} \end{equation}
This should be viewed as the definition of $\ell$.

For the dilaton field we may similarly write
\be
 \phi({\cal R},t,x)= \phi_0 +\frac{nr}{2\ell} + \frac{1}{n}\phi^{(1)}({\cal R},t,x) \,. \label{eq:phiAnsatz}
\ee
Here and below we suppress the indices of spatial coordinates in function arguments for notational simplicity.
Instead of the dilaton correction, we find it convenient to work with the function
\be
 A_\mathrm{s}^{(1)}({\cal R},t,x) =  A^{(1)}({\cal R},t,x) + \frac{2}{3}\phi^{(1)}({\cal R},t,x)
\ee
which is identified as the scale factor of the metric in the string frame (with the transformation between the string and Einstein frames defined as in the five-dimensional picture).

The vacuum solution is given by setting the leading order VEVs to zero, i.e. $m(t,x)=0$ and $p_i(t,x)=0$. If we set also  $\delta V(\phi) = 0$ the next-to-leading order terms vanish in this case, $A^{(1)}({\cal R},t,x) = f^{(1)}({\cal R},t,x) = C_i^{(1)}({\cal R},t,x) = \phi^{(1)}({\cal R},t,x) =0$. This background is a specific exact solution for $\deltaV=0$, the linear dilaton background~\cite{Aharony:1998ub,Bertoldi:2009yi,Gursoy:2010jh,Fichet:2023xbu,Fichet:2023dju}. However, in general also the vacuum solution contains nontrivial contributions to the functions  $A^{(1)}$, $f^{(1)}$ and $C_i^{(1)}$, see Appendices~\ref{app:nlosolution} and~\ref{app:UVasymptotics}.

Note also that the spatial components $g_{ij}$ of the metric are suppressed by a factor of $1/n$ with respect to the other components of the metric as seen from the leading order term $\delta_{ij}/n$ in our Ansatz~\eqref{eq:gijAnsatz}. This implies that we have rescaled the spatial coordinates, $x_k = \sqrt{n}\,z_k$, where $z_k$ are the ``physical'' coordinates. This is also following the conventions of the large-$D$ expansion.

\section{Black hole effective theory} \label{sec:BHET}

In this section, we analyze some of the basic properties of the generalized effective framework for dynamical black holes found in~\cite{Jarvinen:2024vdi}. We first derive the effective equations, discuss their symmetries, and then derive implications for black hole thermodynamics.

\subsection{Effective equations}

Starting from the Ansatz described in the previous section, we expand the Einstein equations as a series in $1/n$ to find the solution. Specifically, we are interested in the solution where sources of the metric and the dilaton are set to the values of the vacuum solutions, and therefore the sources are unaffected by the black hole. At leading order this is already imposed in our Ansatz, e.g.,~\eqref{eq:fAnsatz} and~\eqref{eq:gijAnsatz}, where the boundary metric reduces to Minkowski (taking into account the scaling of $x^k$) as ${\cal R} \to 0$. But we must require that the next-to-leading order variations of the sources, arising from the corrections $A^{(1)}$, $f^{(1)}$, $A_\mathrm{s}^{(1)}$ and $C_i^{(1)}$, also vanish.

Expanding the Einstein equations as a series in $1/n$ while keeping ${\cal R}$ (rather than $r$) fixed, the leading nontrivial  equations readily imply the energy conservation equation
\be
 \pa_t m = \ell \pa_i \pa^i m -\pa_i p^i \,.
\ee
Going to next-to-leading order, we find that the perturbations $f^{(1)}$, $A^{(1)}$, and $A_\mathrm{s}^{(1)}$ satisfy the coupled equations
\begin{align}
\label{eq:A2ndorder}
    3 {\cal R} \pa_{\cal R}^2A^{(1)}({\cal R},t,x)+3 \pa_{\cal R}A^{(1)}({\cal R},t,x)+2 \pa_{\cal R}A_\mathrm{s}^{(1)}({\cal R},t,x)+\frac{p_i(t,x)^2}{2 m(t,x)} & = 0 & \\
      \label{eq:f2ndorder}
  \pa_{\cal R}^2f^{(1)}({\cal R},t,x)-3 m(t,x) \pa_{\cal R}A^{(1)}({\cal R},t,x)-\frac{\ell \pa_ip^ i(t,x)}{{\cal R}}+\frac{1}{2} p_i(t,x)^2 &= 0 &  \\
  {\cal R} \pa_{\cal R}f^{(1)}({\cal R},t,x)+3(2\!-\! {\cal R} m(t,x)) {\cal R}\pa_{\cal R}A^{(1)}({\cal R},t,x)+2(1\!-\!{\cal R}m(t,x)) {\cal R} \pa_{\cal R}A_\mathrm{s}^{(1)}({\cal R},t,x) \!\!\nonumber&&\\
  +2 A_\mathrm{s}^{(1)}({\cal R},t,x)-f^{(1)}({\cal R},t,x)-\ell {\cal R} \pa_ip^i(t,x)
  +\frac{{\cal R} p_i(t,x)^2}{m(t,x)}+\delta V \left(\frac{\log {\cal R}}{2}\right) &= 0 &
  \label{eq:nloconstraint}
\end{align}
Instead of trying to solve these equations directly, one can first carry out integrations with respect to $\cal R$. For example,~\eqref{eq:f2ndorder} can be obviously integrated, and~\eqref{eq:A2ndorder} can also be directly integrated. After carrying out a third, slightly more complicated integration, we find that:
\begin{align}
\label{eq:A1storder}
    3 {\cal R} \pa_{\cal R}A^{(1)}({\cal R},t,x)+2 A_\mathrm{s}^{(1)}({\cal R},t,x)+\frac{p_i(t,x)^2}{2 m(t,x)}{\cal R} & = c_1(t,x) & \\
      \label{eq:f1storder}
  \pa_{\cal R}f^{(1)}({\cal R},t,x)-3 m(t,x) A^{(1)}({\cal R},t,x)-\ell \pa_ip^ i(t,x)\log{\cal R}+\frac{1}{2} p_i(t,x)^2{\cal R} &= c_2(t,x) &  \\
\frac{f^{(1)}({\cal R},t,x)}{{\cal R}}-3 m(t,x) A^{(1)}({\cal R},t,x)+\frac{2(1-m(t,x){\cal R}) A_\mathrm{s}^{(1)}({\cal R},t,x)}{{\cal R}} &&\nonumber\\+ \int_0^{\cal R} \frac{\delta V\left(\frac{1}{2} \log \rs\right)}{\rs^2} \, d\rs -\ell  \pa_ip^i(t,x)\log {\cal R}-\frac{2 c_1(t,x)}{{\cal R}} &= c_3(t,x) \ .&
\label{eq:algebraic}
\end{align}
The more complicated integration that results in the last equation arises by integrating a combination of~\eqref{eq:nloconstraint} and~\eqref{eq:A1storder}, as one can verify by taking the derivative of the result. If the integral in the last equation involving $\delta V$ is divergent at $\rs=0$, one can move the starting point to another value as such changes can be absorbed in $c_3$.

Note that Eqs.~\eqref{eq:A1storder}--\eqref{eq:algebraic} are still nontrivially coupled. However, these equations also imply an equation for the first $\cal R$-derivative of $A_\mathrm{s}^{(1)}$. Interestingly, this equation,
\begin{align}
\label{eq:As1storder}
 2(1- {\cal R} m(t,x))\pa_{\cal R}A_\mathrm{s}^{(1)}({\cal R},t,x) +\int_0^{\cal R} \frac{\delta V'\left(\frac{1}{2} \log \rs\right)}{2 \rs^2} \, d\rs && \nonumber\\+c_2(t,x)-c_3(t,x)-\ell \pa_ip^i(t,x)- m(t,x) c_1(t,x) &= 0 &    
\end{align}
becomes independent of the other fields. Therefore the solution can be obtained by direct integration, and the other fields may be solved sequentially from~\eqref{eq:A1storder} and~\eqref{eq:algebraic}. The explicit solution is given in Appendix~\ref{app:nlosolution}.

However, interesting information may be obtained from these equations without solving them explicitly. The physically interesting solution is the one which remains regular at the horizon. Note that the coefficient of the derivative term in~\eqref{eq:As1storder} vanishes at the horizon ${\cal R} = 1/m(t,x)$ (or more precisely at the leading order location of the horizon). Therefore, in order to avoid a logarithmic singularity of $A_\mathrm{s}^{(1)}$, the other terms in the equation must also vanish at the horizon. This gives the regularity condition
\be \label{eq:regcond}
\int_0^\frac{1}{m(t,x)} \frac{\delta V'\left(\frac{1}{2} \log \rs\right)}{2 \rs^2} \, d\rs +c_2(t,x)-c_3(t,x)-\ell \pa_ip^i(t,x)- m(t,x) c_1(t,x) = 0
\ee
In addition to Eqs.~\eqref{eq:A1storder}--\eqref{eq:algebraic}, the next-to-leading order Einstein equations then imply the equation
\begin{align}
 \pa_t p_i(t,x) &= \ell \pa_j \pa^j p_i(t,x) - \pa_j \left(\frac{p^j(t,x) p_i(t,x)}{m(t,x)}\right) + \frac12 \delta V^\prime\left(-\frac{\log m(t,x)}2\right)\pa_i m(t,x) & \nonumber\\
  \label{eq:effeq2c1}
 &\phantom{=} - m(t,x)\pa_i c_1(t,x) \,,&
\end{align}
where we already used the constraint~\eqref{eq:regcond}. The constraint removes the dependence on $c_2$ and $c_3$, but $c_1$ still remains. However, this remaining function turns out to be a source term (see Appendix~\ref{app:UVasymptotics}). Therefore it should be a constant, which eliminates the last term in~\eqref{eq:effeq2c1}.

The final effective equations governing the dynamics of the black hole solutions are therefore the following~\cite{Emparan:2015gva,Emparan:2016sjk}:
\begin{align}
 \pa_t m &= \ell \pa_i \pa^i m -\pa_i p^i \,,
 \label{eq:effeq1} \\
 \pa_t p_i &= \ell \pa_j \pa^j p_i - \pa_j \left(\frac{p^j p_i}{m}\right) + \frac12 \delta V^\prime\left(-\frac12\log m\right)\pa_i m \,.
  \label{eq:effeq2}
\end{align}

\subsection{Symmetries}

Before we proceed let us comment on the symmetries of the action and how they are reflected in the effective equations. First, since the action does not depend explicitly on any of the coordinates, there is full translation symmetry. That is, given a solution, it remains a solution after an arbitrary shift of the coordinates. Because our Ansatz in~\eqref{eq:AAnsatz}--\eqref{eq:gci} depends explicitly on the holographic coordinate $r$, its translations act on it nontrivially. In order not to affect the large-$n$ counting, we restrict to $\morder{1/n}$ shifts $\delta r$, which means that $\calR$ in~\eqref{eq:calRdef} receives an $\morder{1}$ modification.  We then identify the following transformation
\be
\begin{split}    \label{eq:Rscale}
 &\calR  \mapsto \Lambda_t \calR \ , \quad A_0 \mapsto A_0 + \frac{1}{3}\log \Lambda_t  \ , \quad \phi_0 \mapsto \phi_0 - \frac{1}{2}\log \Lambda_t  \ , \\
&m(t,x) \mapsto \frac{m(t,x)}{\Lambda_t }  \ , \quad p_i(t,x) \mapsto \frac{p_i(t,x)}{\Lambda_t } \,,   
\end{split}
\ee
where $\Lambda_t = e^{n\delta r/\ell}$. The Ansatz is invariant in form under this transformation. Consequently, the next-to-leading Einstein equations~\eqref{eq:A2ndorder}--\eqref{eq:nloconstraint} and the effective equations~\eqref{eq:effeq1}, \eqref{eq:effeq2} are also invariant under the symmetry, provided that we also add a modification of $\deltaV$:
\be
 \deltaV(\phi-\phi_0) \mapsto \deltaV\left(\phi-\phi_0 -\frac{1}{2}\log \Lambda_t\right) \ . 
\ee
This term adds a slight logarithmic adjustment in the terms involving $\delta V$. This is required because, in order to simplify our equations, we included the factor of $\phi_0$ in the definition of $\delta V$ in ~\eqref{eq:potential}, which breaks the invariance of the Ansatz under the symmetry, unless an extra adjustment is made. If we replace $\delta V(\phi-\phi_0)$ in this definition by  $\delta V(\phi)$, such an adjustment is not needed, as all the occurrences of $\deltaV$ in the equations also contain a $\phi_0$-term.

Another symmetry of the action is a scaling symmetry where all components of the metric scale as $g_{\mu\nu} \mapsto \Lambda_s^2g_{\mu\nu}$ compensated by scalings of all the coordinates, $x^\mu \mapsto x^\mu/\Lambda_s$, where $\Lambda_s>0$. This is a symmetry because it leaves the line element invariant. Naively it appears that applying the scaling symmetry to our Ansatz will be complicated because $\calR$ in~\eqref{eq:calRdef} does not transform simply when $r$ is scaled. However, the scaling of the metric components causes $A_0$ to transform, so that $\ell$ which is defined through~\eqref{eq:V0}, also transforms. It turns out that $r/\ell$, and therefore $\calR$, are actually invariant. The full transformation rules are given by
\be \label{eq:scaletr}
 r \mapsto \frac{r}{\Lambda_s} \ , \quad t \mapsto \frac{t}{\Lambda_s} \ , \quad x^i \mapsto \frac{x^i}{\Lambda_s} \ , \qquad 
 A_0 \mapsto A_0 + \log \Lambda_s \ , \quad  \ell  \mapsto \frac{\ell}{\Lambda_s}  \ .
\ee
This transformation is indeed an exact symmetry of the equations~\eqref{eq:A2ndorder}--\eqref{eq:nloconstraint} and~\eqref{eq:effeq1}, \eqref{eq:effeq2}: derivatives with respect to $t$ or $x_i$ are accompanied by extra powers of $\ell$ in these equations, which leads to their invariance.

The metric in~\eqref{eq:gds2} allows for additional isometries, where one scales $t$ and/or $x^i$ independently and compensates by adjusting the various functions in the metric. However, we have fixed those isometries by requiring that $f$ approaches one and $g_{ij}$ approaches $\delta_{ij}/n$ at the boundary, so they do not act on our large-$D$ Ansatz.

Apart from the above symmetries of the solution, where the components of the action do not transform explicitly, let us point out that there may also be symmetries where parts of the action change. An example is a trivial shift symmetry of the dilaton, where one changes also the potential:
\be
 V(\phi) \mapsto V(\phi-\delta \phi) \ , \qquad \phi \mapsto \phi +\delta \phi  \ ,
\ee
where $\delta \phi$ is a constant. This is indeed a symmetry, because the dilaton only appears in the action~\eqref{eq:EDaction} through the kinetic term and the potential. In our Ansatz, this symmetry only affects the value of the constant $\phi_0$ which may be therefore chosen arbitrarily without affecting any physical observables. Another example is a symmetry of the equations of motion, where one scales the metric as $g_{\mu\nu} \mapsto g_{\mu\nu} \Lambda_V$ and the potential as $V \mapsto V/\Lambda_V$. This acts on our Ansatz as
\be
 A_0\mapsto A_0+\frac12\log\Lambda_V \ ,\qquad V_0\mapsto V_0/\Lambda_V \ ,
\ee
leaving everything else invariant.

The system admits yet another symmetry of different kind~\cite{Emparan:2015gva}. As we pointed out in Sec.~\ref{sec:model}, the spatial coordinates $x^i$ we are using here are related to the physical coordinates through a rescaling with a factor of $1/\sqrt{n}$. Therefore, in addition to the presence of finite temperature, Lorentz covariance is broken by this rescaling. Moreover, the gauge choice implied by our Ansatz above breaks Lorentz covariance. However, as it turns out, the effective equations still satisfy a Galilean boost symmetry, given by
\begin{equation}
        t \mapsto t\ , \qquad x^i \mapsto x^i- v^i t\ , \qquad p^i \mapsto p^i- mv^i \ ,
\end{equation}
where $v^i$ is an arbitrary boost velocity.  Indeed, the covariance of the equations~\eqref{eq:effeq1} and~\eqref{eq:effeq2} under this transformation can be checked by direct computation. Note that the equations are not independently boost-covariant: the covariance check of the second equation uses the first equation.
However, our Ansatz in~\eqref{eq:gds2}--\eqref{eq:gci} and the perturbed Einstein equations do not transform simply under such boosts, which apparently is due to the Lorentz-covariance breaking gauge conditions we impose. That is, the gauge invariance of the effective equations explains why they display the boost symmetry in its simple form.
We also note that the Galilean boost symmetry strongly protects the form of the effective equations. We do not see any obvious generalization of~\eqref{eq:effeq1} and~\eqref{eq:effeq2} that would still remain covariant under these boosts.

\subsection{Black hole thermodynamics}
It is possible to write simple formulas for thermodynamic properties of the black hole solutions.

The entropy density in the model is obtained by evaluating the area of the black hole. Dropping subleading terms in $1/n$, it is given by
\begin{equation} S
= \frac{e^{3A_0} }{4G_5} m \,.\label{eq:entropy}
\end{equation}
In this expression, we also dropped a term $1/n^{3/2}$ which reflects the change from rescaled to physical coordinates pointed out at the end of Sec.~\ref{sec:model}. That is,~\eqref{eq:entropy} is the entropy density in physical coordinates.

The temperature can be computed by evaluating the derivative of the blackening factor at the horizon, as:
\begin{equation} T = -\frac{1}{4\pi}\frac{\partial f}{\partial r}\bigg\vert_{r=r_h}\,.\end{equation}
The minus sign in this definition is due to the fact that in our convention, the boundary is at $r=-\infty$, such that $r$ decreases as one moves away from the horizon. Including the $1/n$ corrections, the location of the horizon is determined by the equation
\begin{equation}
	f(\calR_h,t,x) = 1 - m(t,x)\calR_h + \frac1n f^{(1)}(\calR_h,t,x) = 0
\end{equation}
which at leading order is solved by
\begin{equation}
	\calR_h = \frac{1}{m(t,x)}\left(1 + \frac1n f^{(1)}(\frac{1}{m(t,x)},t,x)\right)
\end{equation}
To get to the temperature, we can write:
\begin{equation}
	\frac{\pa f}{\pa r}\bigg\vert_{r=r_h} = \frac{n}{\ell} \calR \frac{\pa f}{\pa \calR} \bigg\vert_{\calR=\calR_h} = -\frac{n}{\ell}\left[1 + \frac1n\left(f^{(1)} - \calR \pa_\calR f^{(1)}\right)\bigg\vert_{\calR = \frac{1}{m(t,x)}}\right]
\end{equation}

We can derive an expression for the temperature directly from the Einstein equations, without an explicit solution for $f^{(1)}$. Namely, subtracting eq.~\eqref{eq:A1storder} from~\eqref{eq:nloconstraint} and evaluating the equation at $\calR = 1/m(t,x)$, taking account of the regularity condition~\eqref{eq:regcond}, yields directly the equation
\begin{equation} \label{eq:temperaturewithp}
\left(f^{(1)} - \calR \pa_\calR f^{(1)}\right)\Big\vert_{\calR = \frac{1}{m(t,x)}} = \deltaV(-\frac12\log m(t,x)) + \frac{p_i(t,x)^2}{2m(t,x)^2} - \frac{\ell \pa_i p^i(t,x)}{m(t,x)} + c_1
\end{equation}
As noted below eq.~\eqref{eq:effeq2c1}, $c_1(t,x)$ is in fact a source term and constant. Since a constant shift does not affect the rest of our discussion, we omit it from expressions below. For static black holes with $m(t,x) = m = Const.$ and $p = 0$, we find that the temperature is given by
\begin{equation} \label{eq:temperature}
	T = \frac{n}{4\pi \ell}\left[1 + \frac1n \delta V\left(-\frac12 \log m\right)\right]\,. \end{equation}
The leading order term is independent of the size of the black hole, which is a feature of the linear dilaton model~\cite{Bertoldi:2009yi}. The first correction is directly related to the potential as
\begin{equation}
	\delta T = \frac{1}{4\pi\ell}\deltaV(-\frac12\log m) \,.
\end{equation}

We can define also other thermodynamic quantities. For example, the speed of sound is given by
\begin{equation} \label{eq:speedofsound}
	c_s^2 \equiv n\times\frac{S}{T}\frac{dT}{dS} = -\frac12\deltaV^\prime(-\frac12\log m)
\end{equation}
Note that this expression is in fact $n$ times the physical speed of sound. We explicitly added the factor of $n$ to simplify notation later in the article. The physical speed of sound is of order $1/n$.

We can also introduce a notion of free energy using:
\begin{equation} \label{eq:freeenergy}
	d F \equiv - S dT = \frac{e^{3A_0} }{32 \pi G_5 \ell} \deltaV^\prime(-\frac12\log m)dm
\end{equation}
Note that unlike previous expressions, which were naturally defined for static solutions where $m$ is constant, the free energy is defined for systems in which we vary $m$. It will be particularly useful for the domain wall and bubble solutions we study in the following.

\subsection{Energy-momentum tensor} \label{sec:Tmunu}

We then show how the energy-momentum tensor is extracted from the solution.
We present here the idea, see Appendix~\ref{app:FG} for details.

In order to compute the energy-momentum tensor we will write the geometry in the Fefferman--Graham coordinates~\cite{deHaro:2000vlm,Skenderis:2002wp}. Recall that the Fefferman--Graham coordinates are those where mixed components between the space-time and holographic coordinates are absent in the metric, and the metric reduced to the holographic component takes a fixed form similar to that of the pure AdS metric. To write the geometry in this form, we first switch from Eddington--Finkelstein time $t$ to regular time coordinate $\hat t$ through
\be
 t = \hat t - \frac{\ell}{n}\log{\cal R}
\ee
and then redefine the coordinates perturbatively in ${\cal R}$ in order to obtain the Fefferman--Graham form.
Details are given in Appendix~\ref{app:FG}.
The leading components (in $1/n$) of the energy momentum tensor, as extracted from the Fefferman--Graham metric, are
\begin{align} \label{eq:Tmunufinal}
  \widehat T_{tt}( t, x) &= %
  m( t, x) \,,\qquad \widehat T_{ti}( t, x) = -p_i( t, x)+\ell \pa_i m( t, x) &\\
 \widehat T_{ij}( t, x) %
&=\frac{1}{2} I( t, x) \, \delta_{ij} + \frac{p_i( t, x)p_j( t, x)}{m( t, x)}-\ell \pa_i p_j( t, x)-\ell \pa_j p_i( t, x) + \ell^2 \pa_i\pa_j m( t, x) &\nonumber
\end{align}
where
\be \label{eq:Idef}
 I(t,x) \equiv \int^{1/m(t,x)}\frac{d\rs}{\rs^2} \delta V'\left(\frac{1}{2} \log \rs\right) \ .
\ee
In this expression we left the lower limit unspecified, because the convenient choice depends on the setting.  %
If the near-boundary behavior of $\deltaV$ is such that the integral is convergent at $u=0$, we can set the lower limit to be zero.
See the discussion in Appendix~\ref{app:UVasymptotics} for the divergent case. However, in other settings, such as the application to IHQCD, it may be convenient to set the limit to $u=\infty$. In IHQCD, with this convention the integral represents the pressure difference between the confined and deconfined phases, because (as we shall see in Sec.~\ref{sec:application}) setting $m=0$ takes one to the confined phase in IHQCD.

Somewhat surprisingly, the expression for  $\widehat T_{ij}$ in~\eqref{eq:Tmunufinal} is slightly different from the result obtained in~\cite{Andrade:2020ilm} by using the Brown--York energy-momentum tensor~\cite{Brown:1992br,Balasubramanian:1999re} (see Appendix~\ref{app:FG} for further discussion). That expression depends on the time derivative of $m$. Even after imposing the effective equation~\eqref{eq:effeq1}, the results differ by a contribution $\propto (\partial_i \partial_j - \delta_{ij}\partial^2)m$. It can be checked that a similar discrepancy also appears in the $D$-dimensional framework, i.e., the results obtained from the Brown--York tensor and the coefficients of the Fefferman--Graham expansion are different already in that case, and it has nothing to do with the reduction to five-dimensions or the generalized potential $\delta V$ which we introduce in this article. Instead, it seems to reflect the fact that the near-boundary and large-$D$ limits do not commute. Note also that the discrepancy does not affect the conservation equations of the tensors, because $\partial^i(\partial_i \partial_j - \delta_{ij}\partial^2)m = 0$. In this article we use the Fefferman--Graham expression, which is more natural, given that we work in the five-dimensional picture and deriving the Brown--York expression solely with the five-dimensional metric is not possible.

The hats in the expressions~\eqref{eq:Tmunufinal} refer to the fact that they are not the components of the physical energy-momentum tensor. In order to read off the components $T_{\mu\nu}$ of the physical tensor we can switch to the unrescaled coordinates $z^k=x^k/\sqrt{n}$. Doing so, we find that they are given by (up to a common trivial overall factor)
\be \label{eq:Tmunuphys}
 T_{tt}(t,z) = \mathcal{N}\widehat T_{tt}(t,x) \,,  \qquad T_{ti}(t,z) = \mathcal{N}\frac{1}{\sqrt{n}}\widehat T_{ti}(t,x) \,, \qquad T_{ij}(t,z) = \mathcal{N}\frac{1}{n}\widehat T_{ij}(t,x) \ ,
\ee
where $\mathcal{N}$ is a proportionality constant depending on the normalization of the action. The value of $\mathcal{N}$ can be obtained by using the dictionary of the high dimensional AdS$_D$ space through the dimensional reduction of Appendix~\ref{app:dimred} (see~\cite{Gursoy:2015nza}), or simply by using the fact that $ T_{tt} \approx ST$ with the expressions~\eqref{eq:entropy} and~\eqref{eq:temperature}.
We find that
\be
 \mathcal{N} = \frac{n\,e^{3A_0}}{16\pi G_5 \ell} \,.
\ee

For the thermodynamics of static black holes, the most important implication from the extraction of $T_{\mu\nu}$ is the expression for the pressure. For static configurations, we obtain the expression
\be \label{eq:pressure}
 P \equiv T_{xx} = \frac{\mathcal{N}}{n}\widehat T_{xx} = \frac{e^{3A_0}}{32\pi G_5 \ell} I
\ee
with $I$ given in~\eqref{eq:Idef}. Differentiating, we obtain
\be
 dP = -\frac{e^{3A_0}}{32\pi G_5 \ell}  \deltaV^\prime(-\frac12\log m)dm
\ee
in precise agreement with the expectation that $P = -F$, with the free energy defined using black hole thermodynamics in~\eqref{eq:freeenergy}.

Therefore, taking into account the rescaling factors from the derivatives with respect to spatial coordinates, the standard conservation equations in physical coordinates have the same form as the conservation equations for $T_{\mu\nu}$ of~\eqref{eq:Tmunufinal} in the rescaled coordinates:
\be
\left\{\begin{array}{cc}
    \frac{\pa}{\pa t}T^{tt} +\frac{\pa}{\pa z^i}T^{it}   & = 0  \\
    \frac{\pa}{\pa t}T^{tj} +\frac{\pa}{\pa z^i}T^{ij} & = 0
\end{array}\right.
\qquad
\Leftrightarrow
\qquad
\left\{\begin{array}{cc}
    \frac{\pa}{\pa t} \widehat T^{tt} +\frac{\pa}{\pa x^i} \widehat T^{it}   &= 0  \\
    \frac{\pa}{\pa t} \widehat T^{tj} +\frac{\pa}{\pa x^i} \widehat T^{ij} & =0
\end{array}\right.
\ee
By inserting the expressions~\eqref{eq:Tmunufinal} to the conservation equations, it is straightforward to verify that the conservation equations yield exactly the two effective equations~\eqref{eq:effeq1} and~\eqref{eq:effeq2}. In this computation, the derivative of the pressure term in~\eqref{eq:Idef} gives the potential-dependent term in~\eqref{eq:effeq2}.

\section{Domain walls} \label{sec:dw}

In this section, we discuss interfaces either between two black hole phases or between a black hole and a horizonless geometry. We start by giving more details about the planar domain walls (which depend only on a single spatial coordinate) considered in~\cite{Jarvinen:2024vdi}. We then discuss spherical or cylindrical bubbles, and finally non-equilibrium steady state walls with nonzero momentum flow.

\subsection{Planar domain walls} \label{sec:dw_planar}

The time-independent solutions of the effective equations~\eqref{eq:effeq1}--\eqref{eq:effeq2} can describe static domain walls. The simplest solutions are in the planar case, where we assume dependence on a single spatial coordinate, which we can choose to be $x_1$. To keep notation simple we will denote it as $x$ below. With the Ansatz:
\begin{equation}
    m = m(x)\,,\qquad p_i = p_i(x)\,,
\end{equation}
the mass conservation equation tells us that
\begin{equation}
    p_1^\prime = \ell m^\dprime
\end{equation}
which we can solve immediately as $p_1(x) = \ell m^\prime(x) + p_0$, where $p_0$ is an integration constant. For now, we consider  only zero momentum solutions where we choose $p_0 = 0$. The transverse components $p_2$ and $p_3$ can also be set to zero. We will return to solutions with non-zero momentum in section~\ref{sec:NESS} below.

Then, substituting $p_1 = \ell m^\prime$ into the remaining effective equation leads to the planar domain wall equation:
\begin{equation} \label{eq:dw_planar}
    m^{(3)} - \frac{2m^\prime m^\dprime}{m} + \frac{(m^\prime)^3}{m^2} = -\frac{1}{2\ell^2}\deltaV^\prime(-\frac12\log m) m^\prime
\end{equation}

We can write general conditions that solutions should satisfy by integrating the equation, and then give a physical interpretation to the conditions using black hole thermodynamics. The equation can be integrated in two ways. By first dividing the equation by $m(x)$ and then integrating w.r.t. $x$ between any two points $x = a$ and $b$, any solution should satisfy that:
\begin{equation}
	\label{eq:dw_int_1}
	\left(\frac{m^{\prime\prime}}{m} - \frac{(m^\prime)^2}{2m^2}\right)\bigg\vert_{x=a}^{b} = \frac{1}{\ell^2}\left(\deltaV\big(-\frac12\log m(b)\big)-\deltaV\big(-\frac12\log m(a)\big)\right)
\end{equation}
A second condition is obtained by integrating equation~\eqref{eq:dw_planar} directly, \textit{i.e.} without first dividing by $m(x)$, with the result that:
\begin{equation} \label{eq:dw_int_2}
	\left(m^\dprime - \frac{(m^\prime)^2}{m} \right)\bigg\vert_{x=a}^{b} = -\frac{1}{2\ell^2} \int_{m(a)}^{m(b)} d\mu \,\deltaV^\prime\left(-\frac12\log \mu\right) 
\end{equation}

For appropriately chosen $\delta V$, the solutions of this system can describe solutions with two domains separated by a domain wall, with $m(x) \approx m_L$ on the left side of the wall ($x \to -\infty$) and $m(x)\approx m_R$ on the right ($x\to\infty$). 

We can show that small perturbations around the constant values of $m$ away from the domain wall vanish exponentially. For domain wall solutions, at $|x| \gg \ell$, we can assume the solutions are given by a small perturbation around the asymptotic value $m_A$:
\begin{equation} \label{eq:dw_asymptotic_expansion}
	m(x)= m_A+ \delta m(x)\,.
\end{equation}
Eqs.~\eqref{eq:dw_int_1} and \eqref{eq:dw_int_2} both imply the linearized equation
\begin{equation} \label{eq:dw_asymptotic_eq}
	\delta m^\dprime +\frac{1}{2\ell^2}\deltaV^\prime\left(-\frac{1}{2}\log m_A\right)\delta m =0
\end{equation}
for $\delta m$, which has the exponential solutions
\begin{equation}
	\delta m (x) \propto e^{\pm\sqrt{-\frac12\deltaV^\prime(-\frac12\log m_A)}x / \ell}\,.
\end{equation}
We should choose the positive (negative) sign for $x$ approaching $-\infty$ ($+\infty$). For a domain wall solution interpolating between $m(x\to-\infty) = m_L$ and $m(x\to\infty) = m_R$, the perturbation at both ends of the domain wall vanishes exponentially with coefficient $\sqrt{-\frac12\deltaV^\prime(-\frac12\log m_{L/R})}$. This last quantity is equal to the speed of sound in each phase, given in eq.~\eqref{eq:speedofsound}.

We can see that the sign of $\deltaV^\prime$ needs to be negative at the endpoints, such that domain wall solutions will have positive $c_s^2$ at both endpoints and decay exponentially to their asymptotic values at $x\to \pm \infty$. This also implies that if we take $a\to-\infty$ and $b\to\infty$ in eqs.~\eqref{eq:dw_int_1}--\eqref{eq:dw_int_2}, the derivative contributions from the left hand sides of the equations vanish, and we are left with:
\begin{equation}
	\label{eq:dw_con1}
	 \frac{1}{\ell^2}\left(\deltaV\big(-\frac12\log m_R\big)-\deltaV\big(-\frac12\log m_L\big)\right) = 0
\end{equation}
\begin{equation} \label{eq:dw_con2}
	\int_{m_L}^{m_R} dm \,\deltaV^\prime\left(-\frac12\log m\right) = 0
\end{equation}

Then, by using the definition of the black hole temperature in eq.~\eqref{eq:temperature} we get that eq.~\eqref{eq:dw_con1} is in fact the condition that:
\begin{equation} \label{eq:dw_deltaT}
	\Delta T \equiv T_R-T_L = 0 \,,
\end{equation}
that is, the temperature is equal in both phases. We should note that eq.~\eqref{eq:temperature} gives the temperature of a solution with constant $m$. The temperatures are understood to be given by
\begin{equation}
	T_{L/R} = \frac{n}{4\pi \ell}\left[1 + \frac1n \delta V\left(-\frac12 \log m_{L/R}\right)\right] \,.
\end{equation}
evaluated for the appropriate asymptotic values of $m$, when we are sufficiently far away from the domain wall and the solutions can be treated as nearly constant.

Likewise, we can see that with our definition of free energy in eq.~\eqref{eq:freeenergy}, the condition that eq.~\eqref{eq:dw_con2} is imposing is that the change in the free energy across the domain wall,
\begin{equation} \label{eq:dw_deltaF}
	\Delta F \equiv F_R - F_L =-\int_{m_{L}}^{m_{R}}\!\! S\,dT = 0
\end{equation}
has to vanish. Note that this is interpreted as the thermodynamics of homogeneous solutions. Here $m$ depends on $x$, while $S$ and $T$ are strictly speaking defined for $x$-independent solutions. While it is possible to interpret the integral over the entire domain wall as the change in free energy, one should not take it too literally at finite values of $x$.

There is one important caveat. In the preceding discussion we implicitly assumed that $m_L$ and $m_R$ are both nonzero. We now discuss the case where we consider a domain wall interpolating between a phase with nonzero $m$ and a phase where $m=0$.
Taking $m \to 0$ in our Ansatz means considering the limit of small black holes, and in Einstein dilaton gravity, typically this limit takes one to the so-called thermal gas solution, which is interpreted as the dual of the confined phase (see, e.g.,~\cite{Gursoy:2008bu,Gursoy:2008za}). Therefore, the domain wall ending at zero $m$ is interpreted to be the domain wall between deconfined and confined phases (see also~\cite{Aharony:2005bm,Figueras:2014lka}).
Arguably, these solutions are more interesting physically and will be our focus in section \ref{sec:application}.

We can repeat the derivation, assuming that $m_L$ is finite and $m_R = 0$. In the integrated equations, we take only the endpoint $a\to-\infty$. Since $m_L$ is finite, the derivative terms vanish at $x\to-\infty$ as argued before. At large positive $x$ we write the asymptotic expansion~\eqref{eq:dw_asymptotic_expansion} now with $m_A = 0$. Eq.~\eqref{eq:dw_int_2}, from which we derived the condition on the free energy, can be expanded in a similar way. One can introduce a formal expansion parameter, writing $m(x) = \epsilon\, \delta m(x)$, and write the equation as a series in $\epsilon$. At zeroth order, we obtain again the condition that $\Delta F = 0$ across the domain wall. At first order we obtain the equation:
\begin{equation}
	\delta m^\dprime - \frac{(\delta m^\prime)^2}{\delta m} +\frac{\deltaV^\prime(\infty)}{2\ell^2}\delta m = 0
\end{equation} 
This is solved by
\begin{equation}
	\delta m = A \exp\left(-\frac{\deltaV^\prime(\infty)}{4}(x/\ell)^2 - B x/\ell \right)
\end{equation}
with arbitrary parameters $A$, $B$. We should require that $\deltaV^\prime(\infty) \geq 0$. Note that by our definition of the speed of sound, this means that we must have that $c_s^2 \leq 0$ as $m\to0$. On the other hand, this does not have any adverse effects in the confined phase where $m=0$.

A last remark is that for some applications it is useful to write the domain wall equation~\eqref{eq:dw_planar} in terms of
\begin{equation}
	\vphi(x) \equiv -\frac12\log m(x)\,,
\end{equation}
by setting $m = e^{-2\vphi}$ in eq.~\eqref{eq:dw_planar}:
\begin{equation}
	\vphi^{(3)} - 2\vphi^\dprime \vphi^\prime + \frac1{2\ell^2}\deltaV^\prime(\vphi)\vphi^\prime = 0
\end{equation}
In particular, expressing the integrated eq.~\eqref{eq:dw_int_1} in terms of $\vphi$ gives
\begin{equation}
	\big((\vphi^\prime)^2 - \vphi^\dprime\big)\big\vert_{x=a}^b = \frac{1}{2\ell^2}\big(\deltaV(\vphi(b)) - \deltaV(\vphi(a))\big)
\end{equation}
which can be useful as it allows one to ``engineer'' a potential $\deltaV$ starting from a particular form of the domain wall solution $m(x)$, in some cases allowing for analytic solutions of the equations.

\subsection{Bubble solutions}
We can consider the planar domain walls as a specific case of a larger class of solutions which also includes cylindrical and spherically symmetric solutions. Before restricting ourselves to the time independent solutions, we start by taking solutions of the form:
\begin{equation} m = m(t,\rho)\,, \qquad p_i = p_i(t,\rho)\end{equation}
where
\begin{equation} \rho = \sqrt{x_1^2+\ldots + x_k^2} \end{equation}
where the cases $k=2$ (cylindrical symmetry) and $k=3$ (spherical symmetry) describe  bubble solutions, and the planar domain wall is given by the choice $k = 1$.\footnote{Strictly speaking, in the planar domain wall case we should set $\rho = x_1 \in (-\infty,\infty)$ rather than $\sqrt{x_1^2} \geq 0$, but for the discussion below the distinction does not matter and equations reduce to the planar case when setting $k=1$.}

It is not hard to see that we can set $p_i = 0$ for all $i \neq 1,\ldots,k$, and that we can keep only the radial component of $p$:
\begin{equation} {\vec p} = p(t,\rho)\, {\hat \rho} \end{equation}
After the change to the spherical coordinates, the effective equations for the two functions describing the dynamic bubble solution, $m(t,\rho)$ and $p(t,\rho)$, are:

\begin{equation}
    \pa_t m = \frac{\ell}{\rho^{k-1}}\frac{\pa}{\pa \rho}\left(\rho^{k-1} \pa_\rho m\right) - \frac{1}{\rho^{k-1}}\frac{\pa}{\pa \rho}\left(\rho^{k-1} p\right)
\end{equation}
\begin{equation}
    \pa_t p = \ell\frac{\pa}{\pa \rho}\left(\frac{1}{\rho^{k-1}}\frac{\pa}{\pa \rho}(\rho^{k-1}p)\right) -\frac{1}{\rho^{k-1}}\frac{\pa}{\pa \rho}\left(\rho^{k-1}\frac{p^2}{m}\right)+\frac12\delta V^\prime\left(-\frac12\log m\right)\pa_\rho m
\end{equation}

For time independent solutions we solve the first equation by setting $p(\rho) = \ell m^\prime(\rho)$ as in the planar case, and the remaining equation for $m(\rho)$ is:
\begin{equation} \frac{\pa}{\pa \rho}\left(\frac{1}{\rho^{k-1}}\frac{\pa}{\pa \rho}(\rho^{k-1}m^\prime)\right) -\frac{1}{\rho^{k-1}}\frac{\pa}{\pa \rho}\left(\rho^{k-1}\frac{(m^\prime)^2}{m}\right)+\frac1{2\ell^2}\delta V^\prime\left(-\frac1{2}\log m\right) m^\prime = 0 \label{eq:dw_spherical} \end{equation}
It is easy to see that setting $k=1$ reduces the equation back to the planar eq.~\eqref{eq:dw_planar}.

As in the planar case, we can integrate the equation in two ways, either directly integrating with respect to $\rho$, or integrating after dividing by $m(\rho)$.

In this way we get the two equations
\begin{equation}
\label{eq:dw_spherical_int_1}
\left(\frac{m^{\prime\prime}}{m} - \frac{(m^\prime)^2}{2m^2} + \frac{k-1}{\rho}\frac{m^\prime}{m}\right)\bigg\vert_{\rho=\rho_1}^{\rho_2} = \frac{1}{\ell^2}\left(\deltaV\big(-\frac12\log m(\rho_2)\big)-\deltaV\big(-\frac12\log m(\rho_1)\big)\right)
\end{equation}
\begin{equation} \label{eq:dw_spherical_int_2}
	\left(m^\dprime - \frac{(m^\prime)^2}{m} + (k-1)\frac{m^\prime}{\rho}\right)\bigg\vert_{\rho=\rho_1}^{\rho_2} \!- (k-1)\int_{\rho_1}^{\rho_2} d\rho  \frac{(m^\prime)^2}{\rho m} = -\frac{1}{2\ell^2} \int_{m(\rho_1)}^{m(\rho_2)} \!dm \,\deltaV^\prime\left(-\frac12\log m\right)
\end{equation}
which are the generalizations of eqs.~\eqref{eq:dw_int_1} and \eqref{eq:dw_int_2}, respectively, the difference being the terms proportional to $k-1$.

As in the planar case, this system can describe solutions with two regions separated by a domain wall, with $m(\rho) \approx m_{\text{in}}$ inside the ``bubble'' ($\rho\to0$) and $m(\rho)\approx m_{\text{out}}$ outside it ($\rho\to\infty$).
However, there will be a clear domain wall where $m(\rho)$ flows from $m(\rho) \approx m_{\text{in}}$ to $m(\rho) \approx m_{\text{out}}$ only if the bubble where $m(\rho) \approx m_{\text{in}}$ is large enough. Specifically, if we take $\rho_1=0$ and $\rho_2=\infty$ in the two equations, the derivative contributions on the left hand side vanish exponentially in $R$, where $R$ is the radius of the bubble, as follows from the exponential asymptotic behavior noted in the previous subsection. In particular, this ensures that the term $m^\prime/\rho$, which we did not have in the planar case, becomes exponentially small in $R$ as $\rho\to0$, and therefore negligible.
In this case, eq.~\eqref{eq:dw_spherical_int_1} implies again that the difference in temperature should vanish:
\begin{equation}
    \Delta T = T_{\text{out}}-T_{\text{in}} = 0 \,,
\end{equation}

The right hand side of eq.~\eqref{eq:dw_spherical_int_2} can again be identified with the change in free energy between the outside and inside of the bubble. Unlike the planar case, in the cylindrical or spherical case this is non-zero, since we have an additional integral term on the left hand side.

To better understand this contribution, we consider a large bubble expansion, as follows. We will assume that $m(\rho)$ is varying only in the region $\rho = R +\delta \rho$, where $R$ is the radius of the bubble, and $\delta \rho\ll R$. Since the scale of the domain wall is $\sim \ell$, which sets also the scale of the variation, we should require that $R \gg \ell$.

We can then consider a systematic expansion of $m$ around the large bubble radius $R$ as:
\begin{equation} \label{eq:largeRexp}
    m(R+\delta \rho)=m_0(\delta \rho)+\frac{m_1(\delta \rho)}{R}+\frac{m_2(\delta \rho)}{R^2}+\ldots
\end{equation}
and expand eq.~\eqref{eq:dw_spherical}. As expected, the zeroth order term $m_0$ should satisfy the planar domain wall equation~\eqref{eq:dw_planar}, since that corresponds to the limit $R\to\infty$. Moving to the next order, we obtain the equation for the leading correction
\begin{equation} \label{eq:m1eq}
    \frac{d}{d\,\delta\rho} \left(m_1^\dprime-\frac{2m_0^\prime}{m_0}m_1^\prime+\frac{(m_0^\prime)^2}{m_0^2}m_1+\frac{1}{2\ell^2}\deltaV^\prime(-\frac12\log m_0)m_1\right) = -(k-1)m_0^\dprime + (k-1)\frac{(m_0^\prime)^2}{m_0}
\end{equation}
which, given the planar solution $m_0$, is an ODE for $m_1$. One can continue to solve sequentially higher order corrections in this expansion. We remark that inserting $m_1 = m_0'$ on the left hand side of~\eqref{eq:m1eq} gives back the original planar equation~\eqref{eq:dw_planar}. That is, $m_1 = m_0'$ is the solution of the homogeneous $m_1$ equation. As seen from the expansion~\eqref{eq:largeRexp}, this solution reflects the freedom of making small $\mathcal{O}(1/R)$ shifts in the value of $R$. Actually a similar freedom is already present at the zeroth order, since the equation for $m_0$ is shift-invariant; this maps to the freedom of making $\mathcal{O}(1)$ shifts in $R$. Such freedom can be fixed by imposing conditions (e.g., $m_1(0)=0$) at each order in the expansion.

Returning to equation~\eqref{eq:dw_spherical_int_2}, we can now write the condition that it imposes on the solution in the large bubble approximation at leading order:\footnote{The nominal range of $\delta\rho$ is from $-R$ to infinity, but given the asymptotic behavior of our solutions, we can safely take the lower bound to $-\infty$.}
\begin{equation} \label{eq:deltaFbubble}
    \frac{16\pi G_5}{\ell e^{3A_0}}\Delta F = \frac{k-1}{R} \int_{-\infty}^\infty d({\delta\rho}) \frac{(m_0^\prime)^2}{m_0}
\end{equation}
where
\begin{equation}
	\Delta F = F_{\text{out}}-F_{\text{in}} \equiv -\int_{m_{\text{in}}}^{m_{\text{out}}} S\, dT
\end{equation}
is defined as before.

For the physical interpretation of eq.~\eqref{eq:deltaFbubble}, we compare it to the standard soap bubble equation, which can be obtained by extremizing
\begin{equation}
    F(R) = -\Delta p\,V_k(R)  + \sigma A_k(R)
\end{equation}
where $\Delta p = (p_{\text{in}}-p_{\text{out}})$ is the pressure differential, $\sigma$ the surface tension, and $V_k$ and $A_k$ are the volume and surface area of the $k$-dimensional bubble of radius $R$, respectively. However, when applying these formulas, one should recall that the radius $R$ is measured in rescaled coordinates (see the discussion at the end of Sec.~\ref{sec:model}), and for the expressions in physical coordinates, one should rescale $R$ by a factor of  $1/\sqrt{n}$.
The volume and the area are therefore related as $A_k = \frac{k\sqrt{n}}{R}V_k$. The extremization $\pa F/\pa R = 0$ leads to the condition on the size of the bubble~\cite{Coleman:1977py}:
\begin{equation} \label{eq:deltap}
    \Delta p = (k-1)\sqrt{n}\frac{\sigma}{R}
\end{equation}

The pressure difference can be identified with the change in free energy, $\Delta p = \Delta F$. From the comparison of eqs.~\eqref{eq:deltaFbubble} and \eqref{eq:deltap}, we can see that
\begin{equation}
    \sigma = \frac{\ell e^{3A_0}}{16\pi G_5\sqrt{n}}\int_{-\infty}^\infty d(\delta\rho) \frac{(m_0^\prime)^2}{m_0}
\end{equation}
is readily interpreted as the surface tension of the domain wall bubble. It is always positive.

The same result for the surface tension could have been obtained directly from the energy-momentum tensor \eqref{eq:Tmunuphys}. Substituting a planar domain wall solution $m = m_0(x)$ and $p_x = \ell m_0^\prime(x)$, and recalling to insert the factor $1/\sqrt{n}$ from the transition to physical coordinates, we get:
\begin{equation} \label{eq:surfacetension}
    \sigma = \frac{1}{\sqrt{n}}\int_{-\infty}^\infty ({T}_{xx}-{T}_{yy}) dx = \frac{{\cal N}}{n^{3/2}} \int_{-\infty}^\infty \frac{p_x^2}{m} dx = \frac{\ell e^{3A_0}}{16\pi G_5\sqrt{n}}\int_{-\infty}^\infty \frac{(m_0^\prime)^2 }{m_0}dx
\end{equation}
which is exactly the expression in the previous equation.

\subsection{Solutions with nonzero momentum flow} \label{sec:NESS}

To obtain eq.~\eqref{eq:dw_planar} for the planar domain wall we have assumed that the momentum $p_i(x)$ goes to zero in the regions where $m(x)$ is constant. Solutions where this is not the case may have an interesting physical interpretation as non-equilibrium steady states (NESS). We introduce this type of solutions and write the conditions they must satisfy.
NESS states with wall structures have been studied in holography in setups that do not include a phase boundary and are therefore somewhat different from ours. Earlier works study an interface between two heat baths~\cite{Bhaseen:2013ypa,Pourhasan:2015bsa,Ecker:2021ukv} and flows in the presence of a defect~\cite{Sonner:2017jcf,Bachas:2021tnp}. However, there are also studies of moving phase boundaries~\cite{Bea:2021zsu,Janik:2022wsx}, which are linked to NESS states via boosts.

Again let us take the Ansatz $m = m(x)$ and $p_i = p_i(x)$. The most general solutions of the mass conservation equation then become
\begin{equation}
	p_1(x) = \ell m^\prime(x) + p_0
\end{equation}
with a non-zero integration constant $p_0$. In this case the domain wall equation becomes
\begin{equation} \label{eq:dw_p0}
	\ell^2 m^{(3)} - \pa_x \left(\frac{(p_0+\ell m^\prime)^2}{m}\right) +\frac12\deltaV^\prime(-\frac12\log m) m^\prime = 0
\end{equation}
In general, we can also consider solutions with a non-trivial transverse momentum $p_2(x)$, which should solve the equation:
\begin{equation}
	\ell p_2^\dprime - \pa_x \left(\frac{(p_0+\ell m^\prime)p_2}{m}\right) = 0
\end{equation}
This last equation has two non-trivial solutions:
\begin{equation}
	p_2(x) = v_2^{(1)} m(x) + v_2^{(2)} m(x) \exp\left(\frac{p_0}{\ell}\int^{x} \frac{d\xi}{m(\xi)}\right)
\end{equation}
The first solution is proportional to $m(x)$ and represents a transverse Galilean boost. As such it can be set to zero by symmetry without affecting the solution. The second solution is divergent for any domain wall where $m(x)$ goes to constant values as $x\to\pm\infty$, so we will not consider it here. Therefore we may set $p_2 = 0$ for the planar domain wall, and focus on the effect of the longitudinal momentum.

We can better understand the role of the non-zero momentum $p_0$ by considering how the conditions imposed on the domain wall, eqs.~\eqref{eq:dw_deltaT} and~\eqref{eq:dw_deltaF}, are modified by it. One condition, obtained by integrating eq.~\eqref{eq:dw_p0} from $x=a$ to $b$, is
\begin{equation} \label{eq:dw_p0_int1}
	\left(\ell^2 m^\dprime - \frac{(p_0 + \ell m^\prime)^2}{m} \right)\bigg\vert_{x=a}^{b} = -\frac{1}{2} \int_{m(a)}^{m(b)} dm \,\deltaV^\prime\left(-\frac12\log m\right) 
\end{equation}
The second condition is the result of dividing eq.~\eqref{eq:dw_p0} by $m(x)$ and then integrating:
\begin{equation} \label{eq:dw_p0_int2}
	\left(\frac{\ell^2 m^\dprime}{m} - \frac{(p_0+\ell m^\prime)^2}{2m^2}\right)\bigg\vert_{x=a}^b -\ell p_0\int_a^b \frac{m^\dprime}{m^2} dx  = \deltaV(-\frac12\log m(b))-\deltaV(-\frac12\log m(a))
\end{equation}

In both equations, when considering the limit of integrating over the entire domain wall by taking $a\to-\infty$, $b\to\infty$, the derivative terms will vanish as before, and we are left with:
\begin{equation} \label{eq:nesscon1}
	p_0^2 \left(\frac{1}{m_L}-\frac{1}{m_R}\right)	= -\frac12 \int_{m_L}^{m_R} dm \,\deltaV^\prime(-\frac12\log m)
\end{equation}
\begin{equation} \label{eq:nesscon2}
	-\ell p_0\int_{-\infty}^{\infty} \frac{m^\dprime}{m^2} dx  = \deltaV(-\frac12\log m_R)-\deltaV(-\frac12\log m_L) + \frac{p_0^2}{2}\left(\frac{1}{m^2_{R}}-\frac{1}{m^2_{L}}\right)
\end{equation}
where $m_{L/R}$ are the asymptotic values of $m(x)$ at $x \to \mp\infty$.

Eq.~\eqref{eq:nesscon1} can also be expressed as
\begin{equation} \label{eq:nesscon1b}
	\int_{m_L}^{m_R} dm \left(\frac12\deltaV^\prime(-\frac12\log m) + \frac{p_0^2}{m^2}\right) = 0
\end{equation}

As usual, we want to understand the conditions from the point of view of the thermodynamics. In this case, asymptotically we have a solution with constant $m$ and $p$. For an ordinary constant solution, one can perform a Galilean boost in order to go to a frame where $p$ vanishes asymptotically. In the domain wall case, on the other hand, since the transformation is $p_i \to p_i + m v_i$ and we have two different values of $m$, there is no frame in which $p$ is zero on both sides, and it is best to work in the original frame where $p$ asymptotes to $p_0$ on both sides of the domain wall. Moreover, if the boost velocity has a nonzero component in the $x$-direction, the domain wall would be moving at a constant velocity after the boost.
Therefore, it is only in the original frame that the domain wall is static, in the sense that it remains in the same location in $x$.

Returning to the derivation of the temperature in eq.~\eqref{eq:temperaturewithp}, we can see that if we have non-zero momentum, and we continue using the definition of the temperature as $\pa f/\pa r$ at the horizon, we get an expression for the ``boosted temperature'' with an additional term
\begin{equation}
	\delta T_\text{eff} = \frac{1}{4\pi\ell}\left[\deltaV(-\frac12\log m) + \frac{p^2}{2m^2}\right]
\end{equation}
In the Eddington--Finkelstein coordinates that we use, the definition of the temperature in terms of the radial coordinate $r$ does not result in a boost invariant quantity. The expression we obtain is compatible with the result of a boost with velocity $v^2 = \frac{1}{n}\frac{p_0^2}{m^2}$ where the temperature transforms as:
\begin{equation}
	T_{\text{eff}} = \gamma T_0 = \frac{T_0}{\sqrt{1-v^2}} = T_0 \left(1 + \frac1n \frac{p_0^2}{2m^2}\right)
\end{equation}
Here $T_0$ is the temperature of a static black hole defined in eq.~\eqref{eq:temperature}. The explicit factor of $1/n$ in the velocity was due to the rescaling of the spatial coordinates. It is the natural choice since the metric was Lorentz-invariant before the rescaling. Note that the equality $T_\text{eff} = \gamma T_0$ is up to leading order $1/n$ corrections. We cannot say from the present computation whether it will continue to hold at higher order.

We can also define an effective free energy as
\begin{equation}
	d F_{\text{eff}} \equiv -S\, d T_{\text{eff}}
\end{equation}
such that the change in free energy is given by
\begin{equation}
	d F_{\text{eff}} \propto  -m d\left(\deltaV(-\frac12\log m) + \frac{p_0^2}{2m^2}\right) =\frac12 \left(\deltaV^\prime(-\frac12\log m) + \frac{2p_0^2}{m^2}\right) dm
\end{equation}
where we omit here the constant prefactor, found in eq.~\eqref{eq:freeenergy}.

With these definitions, we can write the two conditions as
\begin{equation}
	\Delta F_{\text{eff}} = 0
\end{equation}
and %
\begin{equation}
	\Delta T_{\text{eff}} = -\frac{p_0}{2\pi} \int_{-\infty}^\infty \frac{(m^\prime)^2}{m^3} dx
\end{equation}
where we also integrated by parts the integral appearing in~\eqref{eq:nesscon2}. In this form it is clear that the integral is positive. Therefore, independently of the form of the solution we can note that $\Delta T_{\text{eff}}$ is negative if $p_0$ is positive and vice versa: the direction of the momentum is from high to low (effective) temperature.

We can learn more about the allowed solutions by checking the asymptotics of the solution. Taking $m(x) = m_A + \delta m(x)$, in the case of non-zero $p_0$, the linearized fluctuation equation from eq.~\eqref{eq:dw_p0_int2} becomes
\begin{equation}
	\delta m^\dprime-\frac{2p_0}{\ell m_A}\delta m^\prime+\left(\frac{1}{2\ell^2}\deltaV'\left(-\frac{1}{2}\log m_A\right)+\frac{p_0^2}{\ell^2m^2_A}\right)\delta m=0
\end{equation}
which has the solutions
\begin{equation}
\label{eq:fluctsol}
	\delta m (x) \propto e^{\left(\frac{p_0}{m_A}\pm\sqrt{-\frac12\deltaV^\prime(-\frac12\log m_A)}\right)x / \ell}.
\end{equation}

Let us consider the behavior at $x\to\infty$ where $m_A = m_R$. To avoid unstable solutions that diverge at infinity, we should satisfy the inequality
\begin{equation}
	\frac{p_0}{m_R} \pm \mathrm{Re}\sqrt{-\frac12\deltaV^\prime(-\frac12\log m_R)} < 0
\end{equation}
for at least one of the choices of sign. Likewise, we should have on the $x\to-\infty$ side:
\begin{equation}
	\frac{p_0}{m_L} \pm \mathrm{Re}\sqrt{-\frac12\deltaV^\prime(-\frac12\log m_L)} > 0
\end{equation}
for at least one of the choices of sign. As opposed to the $p_0 = 0$ case, we can now allow $c_s^2(m_{L/R}) =-\frac12\deltaV^\prime(-\frac12\log m_{L/R})$ to be negative, since solutions can still decay thanks to $p_0$.

Let us assume without loss of generality that $p_0$ is positive. If it is negative we can use the parity transformation $x\to -x$ to make it positive. This will amount only to reversing the roles of $m_L$ and $m_R$ in the following discussion.

If $p_0>0$, the condition at $x\to-\infty$ can always be satisfied. Now, at $x\to+\infty$, we must have that\footnote{There may also be solutions in the marginal case, $p_0^2/m_R^2 = c_s^2(m_R)$, in which case the fluctuation has power-like decay. However, we do not attempt to cover the marginal cases here.}
\begin{equation}
	0 < \frac{p_0^2}{m_R^2} < c_s^2(m_R)
\end{equation}
otherwise both solutions for $\delta m$ diverge exponentially, implying that there are no stable domain wall solutions. This last condition tells us that the asymptotic fluid velocity, given by $v_R^2 = p_0^2/m_R^2$ must always be less than the speed of sound. We can conclude that this must be the case in at least one of the sides of the domain wall for any stable solution.

The fluid velocity can exceed the speed of sound on the other side, here $m_L$. In the case where $p_0^2/m_L^2 > c_s^2(m_L) > 0$ there is an additional slowly decaying mode with exponent $p_0/(m_L\ell) - \sqrt{c_s^2(m_L)}/\ell$.

To summarize, there are several types of stable solutions. Non-trivial walls
interpolate between  $m_L$ and $m_R$ which satisfy eq.~\eqref{eq:nesscon1b}. The asymptotic behavior is fixed by the exponent of eq.~\eqref{eq:fluctsol}. A wall needs the right end to be in the subsonic region, $c_s^2(m_R)>p_0^2/m_R^2$. The left end then determines the type of solution.
\begin{enumerate}
\item \emph{$p_0=0$ (equilibrium wall).} The wall exists only for the fine-tuned
pair $(m_L,m_R)$ with $T_L=T_R$ and $\Delta F=0$.
\item \emph{$p_0\neq0$, subsonic on both sides.} This is the continuation of the
equilibrium wall. For given $m_L$, it needs one fine-tuned relation between $m_R$ and $p_0$, and the tails are exponential.
\item \emph{$p_0\neq0$, supersonic on the left}, $p_0^2/m_L^2>c_s^2(m_L)>0$. There is no fine-tuning: any $m_R$ in an open interval works.
\item \emph{$p_0\neq0$, $c_s^2(m_L)<0$, oscillations on the left.} There is again no fine-tuning, and the
approach is decaying oscillation, with the amplitude going as $e^{p_0x/(\ell m_L)}$.
\end{enumerate}

Finally, as we remarked above, one can also boost to a frame where instead of being steady, the wall moves, and the fluid in front of the wall is at rest. This configuration is similar to what is expected for walls of expanding bubbles in the context of first order transitions proceeding via bubble nucleation. Note that after the boost, solutions that are supersonic on the upstream (left) side have wall speed exceeding the speed of sound (detonation) while for subsonic solutions the wall speed is below the speed of sound (deflagration)~\cite{Steinhardt:1981ct,Espinosa:2010hh}. Since the downstream flow velocity is less than the speed of sound, only weak deflagrations and strong detonations arise, with the marginal case, where the fluid velocity reaches the speed of sound, corresponding to Jouguet solutions.

\section{Near-boundary completion}\label{sec:UV}

In this section, we initiate the analysis of the near-boundary completion in our framework. The aim is to construct a full dynamical picture for a five-dimensional geometry flowing from an AdS$_5$ fixed point at the UV boundary to a large-$D$ geometry and a horizon in the IR. This can be obtained by matching the dynamical near-horizon solution obtained from the generalized black hole effective theory with dynamical fluctuations in the rest of the geometry (see e.g.~\cite{Suzuki:2015axa} for related analysis in the large-$D$ literature).

We consider a more general setup than above. We still use the Einstein-dilaton action of~\eqref{eq:EDaction} but work with a generic potential rather than explicitly imposing the large-$n$ expansion of~\eqref{eq:potential} everywhere. That is, we will only assume that the large-$n$ expansion is a good approximation to the full potential in a specific region occurring at large values of the dilaton. As the dilaton decreases, $\deltaV$ will grow large and one enters a region where the large-$D$ approach does not work. In standard holography, one assumes that the dilaton eventually flows to an asymptotic AdS${}_5$ regime, where the dilaton approaches a constant value, which can also be $-\infty$ in case of runaway potentials such as those considered in the improved holographic QCD framework~\cite{Gursoy:2007cb,Gursoy:2007er} which we will discuss in Sec.~\ref{sec:application}. However, the presence of the asymptotic AdS${}_5$ region is not required for our analysis here.

For such dynamical black hole geometries, requiring that a simple matching procedure can be established, three regions must be present (see also Fig.~\ref{fig:regions} below):
\begin{itemize}
 \item The near-boundary region. In this region, the geometry is non-trivial and driven by a general dilaton potential $V(\phi)$ which is not well approximated by the form~\eqref{eq:potential}. However, one is far from the dynamical black hole horizon, so the space-time dependence of the geometry can be treated as small perturbations around the static geometry.
 \item The overlap region. In this region, the potential is of the form~\eqref{eq:potential} with a small perturbation term $\propto\deltaV$. However, one is still far from the black hole horizon. Therefore, the space-time dependence can be described both in terms of small perturbations around the static solution or in terms of the (asymptotics of the) general dynamical Ansatz~\eqref{eq:gds2} of the effective framework. In particular, it must be possible to match the two expressions.
 \item The near horizon region. In this region, the space-time dependence is nonlinear, the black hole blackening factor $f$ deviates significantly from one. The dynamics can no longer be described in terms of small fluctuations about a static background, but it can be described in terms of the effective equations~\eqref{eq:effeq1} and~\eqref{eq:effeq2}.
\end{itemize}

Here we will focus on the solutions in the overlap region, discuss how the matching works, and what are the requirements for it to work. We will also recall briefly the general formalism for fluctuations in the near-boundary region. The dynamics in the near-horizon follows the formalism discussed in the earlier sections of this article.

\subsection{Fluctuations in the near-boundary region}

Following the discussion above, far away from the black hole, we assume a background of the form, in the Eddington--Finkelstein coordinates:
\begin{equation} ds^2 = e^{2A(r)}\left(-dt^2 - 2dt\,dr + \delta_{ij}dz^i dz^j\right) \end{equation}
This is nearly the same as our Ansatz for the black hole solution in eq. \eqref{eq:gds2}. There are two differences. Here we have set the blackening factor $f$ to one since we are considering the region far from the horizon. In addition, here we are using the spatial coordinates $z_i$ which are related to the $x_i$ by the rescaling, $z_i = \frac{1}{\sqrt{n}}x_i$.

The degrees of freedom are the scale factor $A(r)$ and the dilaton $\phi(r)$. For a general dilaton potential $V(\phi)$, the solution for the background satisfies the equations:\footnote{A detailed treatment of this model can be found in  \cite{Kiritsis:2006ua}. We follow this reference closely, although our conventions vary.}
\begin{equation} \label{eq:bgeqs}
    \frac49\left(\phi^\prime\right)^2 = (A^\prime)^2-A^\dprime\,,\qquad e^{2A} V(\phi) = 9(A^\prime)^2+3A^\dprime \,.
\end{equation}

We consider linear fluctuations around the solution of these equations, defining
    \begin{equation} \delta g_{\mu\nu} = e^{2A}\,h_{\mu\nu}(r) e^{-i(\omega t - k z^1)}\,,\qquad
     \delta \phi = \psi(r) e^{-i(\omega t - k z^1)}\,\end{equation}
At this stage we choose the direction of fluctuations to be $z^1$, as there is no preferred direction.

It is convenient to define the functions
\begin{equation}
    \hat{h}_{\mu\nu}(r) \equiv e^{i\omega r}h_{\mu\nu}(r) \,, \qquad \hat{\psi}(r) \equiv e^{i\omega r}\psi(r)
\end{equation}
which essentially apply the transformations from the EF time-coordinate $t$ to the ordinary time, $\hat t = t + r$.

There are two types of fluctuation modes. One is that of the spin-2 fluctuation modes of the metric, and another for the scalar fluctuations, which mix the spin-0 component in $\hat{h}_{\mu\nu}$ with $\hat{\psi}$. The equations for the two types of modes are \cite{Kiritsis:2006ua}
\begin{equation}
        \hat{F}_2^\dprime + 3 A^\prime(r) \hat{F}_2^\prime - (k^2-\omega^2)\hat F_2 = 0
\end{equation}
where $\hat F_2$ is any of the spin-2 modes, and
\begin{equation}
        \hat{F}_0^\dprime + \left(3 A^\prime(r)+2\frac{\zeta^\prime}{\zeta}\right) \hat{F}_0^\prime - (k^2-\omega^2)\hat F_0 = 0
\end{equation}
where $\hat F_0$ is a scalar fluctuation mode. For the last equation we defined the function
\begin{equation} \label{eq:zr}
    \zeta(r) \equiv \frac{\phi^\prime}{A^\prime} \,.
\end{equation}

\subsection{Matching with the large-$D$ expansion}

In order to discuss matching, we first find the solution to the background in the different regions. In the near-boundary region, the dilaton potential drives a nontrivial flow of the dilaton, and the solution can only be constructed numerically. We assume that the flow is characterized by a single energy scale, and choosing the units appropriately the conformal coordinate (which we denote for now $\tilde r$) varies by $\mathcal{O}(1)$ along the flow.

\begin{figure}
\centering
\includegraphics[width=0.8\textwidth]{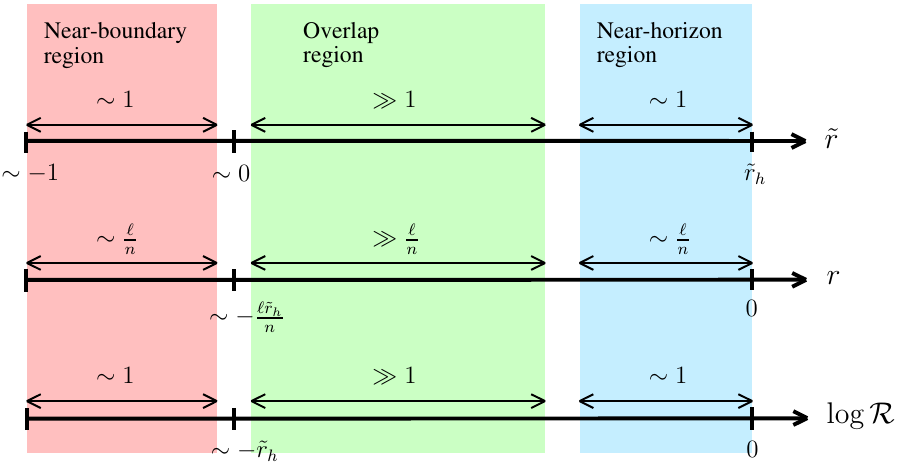}
\caption{Illustration of the three different regions of the holographic flow defined in the text and how they are related to different coordinates, i.e., $\tilde r$, $r$, and ${\cal R}$. Above each coordinate axis, we mark the length of the regions in the corresponding coordinates, while under the axis we mark the values of the coordinates at different locations.}
    \label{fig:regions}
\end{figure}

As $\tilde r$ grows, one enters the overlap region where the potential takes the form of~\eqref{eq:potential}. Without loss of generality, we can take the transition to the overlap region to be located at $\tilde r \approx 0$ (see Fig.~\ref{fig:regions}). Here it is enough to consider the leading order background, i.e., we can neglect all corrections suppressed by powers of $1/n$. Moreover, the overlap region is still far from the horizon, so the blackening factor may be set to one. Then, the solution is the linear dilaton background. Smooth evolution of the background from the near-boundary to the overlap region requires that $A(\tilde r\approx 0) = \mathcal{O}(1)$ and $A'(\tilde r\approx 0) = \mathcal{O}(1)$, where the orders refer to the limit $n \to \infty$. The dilaton should also be continuous at this point. Consequently, the solution is fixed to be
\begin{equation} \label{eq:lineardilatontilde}
    A({\tilde r}) =  \log\frac{\ell}{n}-\frac{\tilde r}{3}\,, \qquad \phi({\tilde r}) = \tilde \phi_0 + \frac{\tilde r}{2}
\end{equation}
where $\tilde \phi_0$ is the value of the dilaton at $\tilde r \approx 0$. Here the values of the coefficients of the linear terms pin down the definition of the coordinate $\tilde r$. This solves~\eqref{eq:bgeqs} if $V(\phi) = \frac{n^2}{\ell^2} e^{\frac{4}{3}(\phi-\tilde \phi_0)}$, similarly to conventions in Sec.~\ref{sec:model}.

Note that the solution only satisfies the condition $A(\tilde r\approx 0) = \mathcal{O}(1)$ up to a term which is logarithmic in $n$. The logarithmic term appears because starting from a potential that is $\mathcal{O}(1)$ near the boundary and strictly following the matching procedure it is not possible to generate the factor $n^2/\ell^2$ in the potential of~\eqref{eq:potential}. %
This appears to be an issue for the UV completion. However, thanks to the exponential dependence on the dilaton of the potential~\eqref{eq:potential} only a logarithmic violation of the matching conditions is required. Moreover, we will consider examples in Sec.~\ref{sec:application} where (despite apparent absence of a parametrically large overall coefficient in the potential), the proportionality factor of the potential appears to be numerically large.

Let us then discuss the boundary between the overlap and the near-horizon regions. Since the size of the black hole is a property of the solution, its location is basically a free parameter. Even if the size of the black hole evolves with space and time, we can define at least locally a characteristic horizon location $\tilde r_h$ around which the horizon fluctuates. Now, in order for the overlap region to be long enough for the matching to work, we must require $\tilde r_h \gg 1$. The condition $\frac{1}{n}\deltaV \ll 1$ must also continue to hold in the whole region, until one reaches the horizon.  A natural scaling property which ensures matching is to require that $\tilde r_h \sim n$. However, this assumption is not required, matching will work so long as $\tilde r_h \gg 1$, regardless of the exact scaling behavior. This is also indicated in Fig.~\ref{fig:regions}. However, one may not want to choose too large $\tilde r_h$, e.g., $\tilde r_h \gg n$, because $\deltaV/n$ should remain small over the overlap region and this may be difficult to achieve for large values of $\tilde r_h$.

Instead of using $\tilde r$, it is convenient to switch to the coordinate $r$ which is natural in the description of the near-horizon dynamics and matches with the notation in the rest of the article. The coordinates are related by a symmetry transformation: we note that the background versions of the shift and scale transformations~\eqref{eq:Rscale} and~\eqref{eq:scaletr} are given by
\be
 A \mapsto A - \log \Lambda \ , \qquad r \mapsto \Lambda\, r + \delta r \ ,
\ee
which indeed leaves the equations~\eqref{eq:bgeqs} invariant.
The specific transformation in this case is (see again Fig.~\ref{fig:regions})
\be
 r = \frac{\ell}{n}\left(\tilde r -\tilde r_h\right) \ ,
\ee
which takes the solution~\eqref{eq:lineardilatontilde} to
\begin{equation} \label{eq:lindil}
    A(r) = A_0-\frac{n r}{3\ell}\,, \qquad \phi(r) = \phi_0 + \frac{n r}{2\ell}
\end{equation}
where $A_0 = - \tilde r_h/3$ and $\phi_0 = \tilde \phi_0 + \tilde r_h/2$. The appearance of the large factors $\propto \tilde r_h$ in the constants $A_0$ and $\phi_0$ may seem worrisome. However, these factors exactly cancel in the effective description. This is seen, for example from the definition of $\ell$: the potential is now $V(\phi) = \frac{n^2}{\ell^2} e^{\frac{4}{3}(\phi-\tilde \phi_0)} = V_0 e^{\frac{4}{3}(\phi-\phi_0)}$ with $V_0 = \frac
{n^2}{\ell^2}e^{\frac 43 (\phi_0-\tilde \phi_0)}$. This matches with the expression in~\eqref{eq:V0} after inserting the formulas of $A_0$ and $\phi_0$, so that the factors involving $\tilde r_h$ cancel.

In conclusion, as also shown in Fig.~\ref{fig:regions}, the overlap region is found when
\be
 - \tilde r_h \ll \frac{n r}{\ell} \ll 0\,,  \qquad e^{-\tilde r_h} \ll {\cal R} \ll 1 \,
\ee
in terms of the coordinates used in the rest of the article.

To match with the large-$D$ expansion, we can consider the fluctuations around the linear dilaton background~\eqref{eq:lindil}.
In this case $\zeta(r)$ defined in eq.~\eqref{eq:zr} is a constant, and so all the fluctuations satisfy the same simple equation:
\begin{equation}
    \hat F^\dprime - \frac{n}{\ell}\hat F^\prime - (k^2-\omega^2)\hat F = 0
\end{equation}
The general solution for the fluctuation modes in the EF coordinate, $F = e^{-i\omega r}\hat{F}$, is then given by
\begin{equation}
    F(r) = e^{-i\omega r}\left(C_+ e^{\alpha_+ r} + C_ -e^{\alpha_- r} \right)
\end{equation}
with
\begin{equation}
\alpha_\pm(\omega,k) = \frac{n\pm\sqrt{n^2+4\ell^2(k^2-\omega^2)}}{2\ell} \label{eq:fluctuationalpha}
\end{equation}
and with the coefficients $C_\pm$ also depending on the frequency and momentum, $C_\pm = C_\pm(\omega,k)$.

So far we have made no assumptions on the values of $k$, or $\omega$. Nor did we have to assume that $n$ is large. 

Now we take that $n\gg 1$. To match with the large-$D$ expansion where we add the $1/n$ corrections to the potential, we have to be in the overlap region, where ${\cal R} = e^{\frac{nr}{\ell}}$ is finite but small, as discussed above.

For the fluctuation modes, we have two choices. The first is to take both $k$ and $\omega$ to be of ${\cal O}(1)$. Then we find that
\begin{equation}
\alpha_+ =\frac{n}{\ell} + \frac{\ell}{n}(k^2-\omega^2) ,\qquad \alpha_- = -\frac{\ell}{n}(k^2-\omega^2)
\end{equation}
Note that while $\alpha_+ \sim n$ at leading order, the product $\alpha_+ r$ is ${\cal O}(1)$ in the region of finite ${\cal R} = e^{\frac{nr}{\ell}}$. In this case, we have the solution
\begin{equation} \label{eq:fluctuationsolution}
    F_{\omega,k}(\calR) = \left(C_+ \calR + C_-\right)(1 -\frac1n i\omega\ell \log\calR)
    \end{equation}
where the leading correction in $1/n$ is the logarithmic correction coming from the factor $e^{-i\omega r} = 1 -\frac{i\omega \ell}{n}\log{\calR}$.

The second option\footnote{Note that both options take one away from the region of nonhydrodynamic modes, which occur at $\omega \sim T \sim n/\ell$ in this setup~\cite{Emparan:2014aba,Emparan:2015rva}. This is consistent with the fact that the near-horizon description only contains hydrodynamic modes.} is to have modes with $\omega$ of order ${\cal O}(1)$, and $k \sim \sqrt{n}$. We denote $k = \sqrt{n} \kappa$ in this limit, with $\kappa \sim{\cal O}(1)$. Recall that we have defined $k$ in terms of the spatial coordinates $z_i$, which are related to the rescaled coordinates we have used in the large-$D$ expansion by $z_i = x_i/\sqrt{n}$. It follows that $k z_i = \kappa x_i$. Since the natural choice of spatial coordinates in the near-horizon region are the rescaled $x_i$, taking $k \sim {\cal O}(\sqrt n)$ is the preferred choice for the matching.

In this case we get an additional correction of ${\cal O}({\frac1n})$ in the solution:
\begin{equation}
	\alpha_+ = \frac{n}{\ell} + \kappa^2 \ell - \frac{\ell}{n}(\kappa^4\ell^2+\omega^2)
\,, \qquad \alpha_- = -\kappa^2 \ell + \frac{\ell}{n}(\kappa^4\ell^2+\omega^2)
\end{equation}
With $k = \sqrt{n}\kappa$, the solution is given by:
\begin{equation} \label{eq:fluctuationsolution2}
    F_{\omega,\sqrt{n}\kappa}(\calR) = C_+\calR+C_- - \frac1n\left[C_+\calR(i\omega\ell-\kappa^2\ell^2)+C_-(i\omega\ell+\kappa^2\ell^2)\right]\log\calR
\end{equation}

So far we did not use any information about how the fluctuations behave in the near-boundary region. As above for the background, the fluctuation solution needs to behave smoothly as we transition from the near-boundary to the overlap region. Assuming that the near-boundary region has nontrivial dilaton flow, the fluctuation solution will have generic source and VEV terms, which are of the same order, as one enters the overlap region.\footnote{The most natural solution to consider is a flow with an AdS boundary, and fluctuation which only has normalizable VEV term at the boundary. However, the flow of the dilaton mixes the source and the VEV, so that when one enters the overlap region, the source and the VEV are of the same size.} Imposing this for the fluctuation solutions at large $n$ gives
\be
 \frac{C_-}{C_+} \sim e^{-\tilde r_h} \ll 1 \ ,
\ee
where the size estimate follows because we take the overlap region to be sizeable, $\tilde r_h \gg 1$. That is, we can neglect the $C_-$ terms when matching.

It is instructive to consider in particular the matching of $F(\calR)$ to the solution for the blackening factor,
\begin{equation} \label{eq:fexpansion}
	f(\calR,t,x) = 1 - m(t,x)\calR + \frac1n f^{(1)}(\calR,t,x)
\end{equation}
where the asymptotic form of the sub-leading piece is:
\begin{equation} \label{eq:fexpansionasym}
	f^{(1)}({\cal R},t,x) = \ell \pa_ip^i(t,x){\cal R}\log{\cal R} +\left[c_2(t,x)-\ell \pa_ip^i(t,x)\right]{\cal R}+ \mathcal{O}\left({\cal R}^2\right)
\end{equation}
The full solution and details on the asymptotic expansions are given in Appendix~\ref{app:UVasymptotics}.

In this case, we see that $C_+$ is sourced directly by $m(t,x)$, as can be seen from the zeroth order term. In Fourier space we can write that $C_+ = -m(\omega,\kappa)$. In the subleading piece, $c_2(t,x)$ reflects the freedom to redefine the mass $m(t,x)$ at subleading order in $1/n$. We do not attempt to match the subleading pieces with the near-boundary fluctuations here.

At the order $1/n$, we can compare the $\calR\log\calR$ terms in the two equations, and see that they also match, since, by the effective equations,
\begin{equation} \ell \pa_i p^i(t,x) = -\ell \pa_t m + \ell^2 \pa_i \pa^i m \,, \end{equation}
which means that in Fourier space we can take
\begin{equation}
	\ell \pa_i p^i(t,x) \to (i\omega \ell - \kappa^2 \ell^2) m(\omega,\kappa)
\end{equation}
which is precisely what we have in eq.~\eqref{eq:fluctuationsolution2} after the identification $C_+ = -m(\omega,\kappa)$ from the leading order solution.

As for the other parts of the geometry, the matching works the same way: as we can see from the expansions in Appendices~\ref{app:UVasymptotics} and~\ref{app:FG}, all the VEV terms in the $1/n$ geometry, not just the $m(t,x)$-term in~\eqref{eq:fexpansion}, have the behavior $\propto {\cal R}$ at small ${\cal R}$, and can be matched with $C_+$ term of~\eqref{eq:fluctuationsolution2} for the various components of the near-boundary fluctuations. After this, computing the final near-boundary VEVs from the fluctuations requires solving them in the near-boundary region. The solutions depend on the choice of potential $V(\phi)$, and to obtain the solutions one typically needs numerical methods. We do not attempt to do this here.

However, note that the expression eq.~\eqref{eq:fexpansionasym} is the asymptotic solution provided that
$\deltaV$ vanishes fast enough in the UV (see Appendix~\ref{app:UVasymptotics} for details and the more general expression). Naively it looks like %
the matching procedure only makes sense for this specific class of potentials. However, this is not the case: the terms involving $\deltaV$ in the expansions, which may be leading with respect to the VEV terms depending on the UV-behavior of $\deltaV$, are independent of the space-time coordinates. They are therefore present even for solutions that do not have any horizon. They should be interpreted as subleading corrections to the linear dilaton solution of~\eqref{eq:lindil}, which are irrelevant for the matching we discuss here. It is the space-time dependent VEV terms, which need to be matched with the space-time-dependent fluctuations, as we have done above, regardless of the near-boundary behavior of the potential. Another way to see this is to note that the space-time independent sources  arising from the terms involving $\deltaV$ are localized at zero frequency and momentum, and therefore essentially decoupled from the matching of the space-time dependent VEVs which generically involve nonzero frequencies and momenta.

\clearpage
\section{Application to improved holographic QCD} \label{sec:application}

In this section, we discuss applications of the formalism to holographic models of the Yang--Mills theory. Specifically we consider the improved holographic QCD model (IHQCD), which means choosing the dilaton potential in~\eqref{eq:EDaction} in a specific manner~\cite{Gursoy:2007cb,Gursoy:2007er,Gursoy:2009jd}. Here we will first discuss how the potential is chosen, and then consider domain wall solutions. Finally we compare our predictions with lattice results for Yang--Mills theory.

\subsection{Potentials and thermodynamics}

We consider the following three choices of the potential:
\begin{align}
\label{eq:VIdef}
V_{\mathrm{I}}(\phi) &= 12\,\biggl[1+V_1 e^\phi+{V_2e^{2\phi}
\over 1+c_\mathrm{IR}e^\phi}+V_\mathrm{IR} \exp\left(-\frac{1}{c_\mathrm{IR}e^{\phi}}\right)
e^{4\phi/3}c_\mathrm{IR}^{4/3}\sqrt{\log(1\!+\!c_\mathrm{IR}e^\phi)}\,\biggr]\,, \\ %
\label{eq:VIIdef}
V_{\mathrm{II}}(\phi) &= 12\,\biggl[\frac{1}{1+\frac{c_\mathrm{UV}}{e^{\phi}}}+V_1 e^\phi+{V_2e^{2\phi}
\over 1+c_\mathrm{IR}e^\phi}+V_\mathrm{IR} \exp\left(\!-\frac{1}{c_\mathrm{IR}e^{\phi}}\right)
e^{4\phi/3}c_\mathrm{IR}^{4/3}\sqrt{\log(1\!+\!c_\mathrm{IR}e^\phi)}\,\biggr]\,, \\
V_{\mathrm{III}}(\phi) &=V_n e^{4\phi/3} \times \left\{
                \begin{array}[2]{cc}
                     e^{-2\phi/3} + c_1 e^{-\phi/3} + c_2 \ ,\qquad   &(\phi \le 0) \\
                      (1+c_1+c_2)\sqrt{1+\phi/\phi_c} \ , \qquad &(\phi \ge 0)
                 \end{array}
\right. \label{eq:VIIIdef}
\end{align}

Here the first two potentials were chosen following~\cite{Jokela:2018ers,Jarvinen:2024vdi} and the third is introduced in this work. The values of the parameters of the first two potentials are
\begin{equation}
 V_1 \approx 1.720 \ , \quad V_2 \approx 1.764 \ , \quad  c_\mathrm{IR} \approx 1.583 \ , \quad V_\mathrm{IR} = 2.05 \,, \quad c_\mathrm{UV} \approx 0.1421 \ .
\end{equation}
The value $c_\mathrm{UV}$ follows~\cite{Jarvinen:2024vdi}, while the other parameters were fitted to the running of the Yang--Mills coupling~\cite{Gursoy:2007cb,Gursoy:2007er} and lattice data~\cite{Panero:2009tv} for the thermodynamics of the Yang--Mills theory~\cite{Jokela:2018ers}. However, the numerical values of the parameters differ from these references because we in addition used the dilaton shift symmetry of the Einstein-dilaton gravity to set the minimum of $V_{\mathrm{I}}(\phi)e^{-4\phi/3}$ to be at $\phi=0$. This amounts to mapping $\phi \mapsto \phi + \Delta \phi$ with $\Delta \phi \approx -0.6393$ with respect to the potentials given in~\cite{Jokela:2018ers,Jarvinen:2024vdi}.\footnote{The shift in principle affects the holographic dictionary of the dilaton field, but since we are focusing here on IR behavior, the precise dictionary is not needed.} In more detail, the potential $V_\mathrm{I}$ is (apart from the shift) that fitted to lattice data in~\cite{Jokela:2018ers} in the full IHQCD model. The modification at small values of $\phi$ in $V_\mathrm{II}$ due to the first term in the square brackets was introduced~\cite{Jarvinen:2024vdi} so that the large-$D$ thermodynamics agrees better with lattice data.

The parameters of the third potential in~\eqref{eq:VIIIdef} are given by
\begin{equation}
 V_n \approx 52.70 \ , \quad c_1 \approx -0.6813 \ , \quad  c_2 \approx 0.8331 \ , \quad \phi_c = 1.778  \ .
\end{equation}
Here $V_n$ was chosen such that the normalization is similar to $V_\mathrm{I}$ and $V_\mathrm{II}$, while the other parameters were fitted to lattice data using the large-$D$ thermodynamics as we will discuss below. The form of the Ansatz for $V_\mathrm{III}$ for $\phi<0$ was chosen to be a simple function of $e^{\phi}$ in the same way as the other potentials. However, the asymptotics at $\phi \to -\infty$ is different: we have $V_\mathrm{III} \sim e^{2\phi/3}$, which was chosen because this produces thermodynamics similar to free theory in the high temperature limit when large-$D$ thermodynamics is imposed, making fitting of the lattice data simpler. In the region $\phi>0$ the potential is weakly constrained by lattice data. The Ansatz $\propto \sqrt{\phi+\phi_c}$ was chosen to mimic the asymptotics found in IHQCD~\cite{Gursoy:2007er}. Notice that the normalization of this part of the potential was determined such that the potential is continuous at $\phi=0$.

These three choices of potentials serve slightly different purposes. As in~\cite{Jarvinen:2024vdi}, the modification of $V_\mathrm{II}$ with respect to $V_\mathrm{I}$ was done such that it improves the matching with lattice\footnote{We will discuss the comparison to lattice data in more detail below.} when one uses the large-$D$ thermodynamics defined in this article (rather than the holographic thermodynamics of full IHQCD). Therefore, the difference between these two potentials directly probes the large-$D$ truncation error in the regime of the potentials constrained by lattice data.
However, the modification between $V_\mathrm{III}$ and $V_\mathrm{II}$ is more drastic. While both the potentials fit the lattice data well using the large-$D$ thermodynamics, the modification is not restricted to the regime (of relatively low $\phi$) where the potentials are best constrained by lattice data, but extends to all values of $\phi$. %
That is, the differences between these two potentials are not a measure of the large-$D$ truncation error, but a measure of the reliability of the holographic approach in general.
Working with such two rather different Ans\"atze in full IHQCD, even if fitted to data in the same way, would also lead to differences in results at the same level as we observe below.

\begin{figure}
\centering
\includegraphics[width=0.7\textwidth]{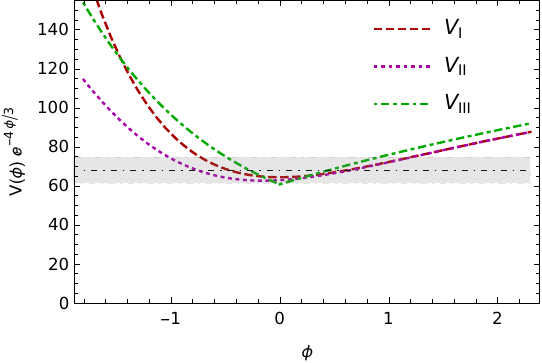}
\caption{The combination $V(\phi)e^{-4\phi/3}$ for the three potential choices considered in this article. Potentials $V_\mathrm{I}$, $V_\mathrm{II}$, and $V_\mathrm{III}$ are shown as the dashed red, dotted magenta, and dot-dashed green curves, respectively. The thin dot-dashed horizontal line shows the constant piece in the potential in the large-$D$ formalism, given by $\hat V_0=68$, while the gray band shows 10\% variation around this value, i.e., the range where $\delta V$ is small.}
    \label{fig:deltaV}
\end{figure}

In order to define the large-$D$ thermodynamics, we then write the potentials as
\begin{equation} \label{eq:deltaVKdef}
 V_K(\phi)e^{-4\phi/3} =\hat V_0 \left[1+\delta V_K(\phi)\right] \ , \qquad K = \mathrm{I}\ ,\ \mathrm{II}\ ,\ \mathrm{III}\ ,
\end{equation}
and apply our formalism using the functions $\delta V_K$ defined by this equation. We choose $\hat V_0 = 68$. The combinations $V(\phi)e^{-4\phi/3}$ for all three potentials are shown in Fig.~\ref{fig:deltaV}.

Notice that because the definition~\eqref{eq:deltaVKdef} does not include explicitly an expansion parameter $n$, the perturbation $\delta V$ should be small and the parameter $\hat V_0$ (which includes a factor of $n^2$, see~\eqref{eq:V0}) should be large. As we see from Fig.~\ref{fig:deltaV},  $\hat V_0$ is indeed rather large and $\deltaV$ is rather small, but only within a limited range of $\phi$: When the colored curves stay within the gray band, $|\deltaV|$ is less than $0.1$. However, even if this range is narrow, it turns out to be the most important region for the description of nontrivial black hole configurations. In order to demonstrate this, we consider below a static domain wall solution between the black hole (deconfined phase) and empty space solution ($m=0$, confined phase).

\begin{figure}\centering
\includegraphics[width=0.45\textwidth]{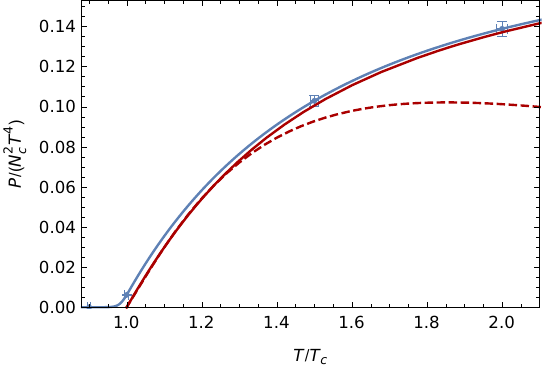}\hspace{5mm}%
\includegraphics[width=0.45\textwidth]{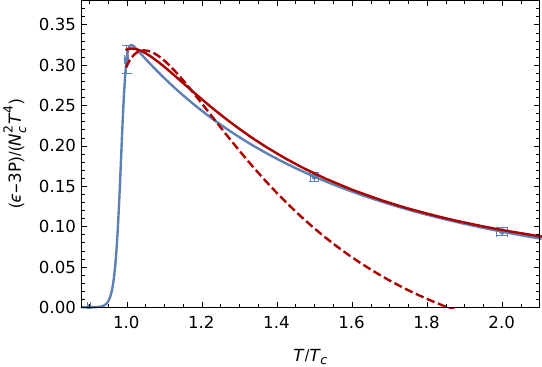}
    \caption{Comparison of the holographic thermodynamics with the potential $V_\mathrm{I}$ and lattice data. The  left (right) panel shows the normalized pressure (interaction measure) as a function of the temperature. The blue curves and error bars show the lattice results for the  thermodynamics of pure Yang--Mills extrapolated to $N_c=\infty$~\cite{Panero:2009tv}. The solid and dashed red curves show the holographic result using the full Einstein-dilaton gravity thermodynamics and the large-$D$ thermodynamics, respectively.}
    \label{fig:lattice_fit_ihqcd}
\end{figure}

However, before going to the domain walls, let us discuss the comparison of the model to Yang--Mills thermodynamics. We start with the potential $V_\mathrm{I}$ given in~\eqref{eq:VIdef}. The fit is shown in Fig.~\ref{fig:lattice_fit_ihqcd}. Here the solid red curve shows the fit to lattice data from~\cite{Jokela:2018ers} using the full Einstein-dilaton gravity holographic dictionary. The dashed red curve is our result, arising from the large-$D$ thermodynamics defined via the equations~\eqref{eq:entropy} and~\eqref{eq:temperature}. The left panel shows the normalized pressure $P/(N_c^2T^4)$ and the right panel shows the normalized interaction measure $(\epsilon-3P)/(N_c^2T^4)$, where $\epsilon$ denotes the energy density.
Notice that since our definitions above do not introduce an expansion parameter $n$, we are effectively setting $n=1$. Moreover, we are fitting only the functional form of the potential, not the overall proportionality constant $e^{3A_0}/G_5$ in the entropy: this constant is separately fitted to data for each fit. Since we are normalizing the temperature to the critical temperature between the $m=0$ (confined) and $m\ne 0$ (deconfined) phases, the normalization of the temperature in~\eqref{eq:temperature} does not enter the analysis either.

We see from Fig.~\ref{fig:lattice_fit_ihqcd} that the large-$D$ thermodynamics agrees well with lattice data and the full Einstein-dilaton fit only near $T = T_c$. This is not surprising because, as seen from~\eqref{eq:temperature}, deviation from $T=T_c$ necessarily means that the subleading term $\propto \deltaV$ becomes large, indicating the breakdown of our framework. However, we stress that this is not an issue for us, because we are mostly interested in the behavior of black holes at temperatures near the phase transition.

\begin{figure}\centering
\includegraphics[width=0.45\textwidth]{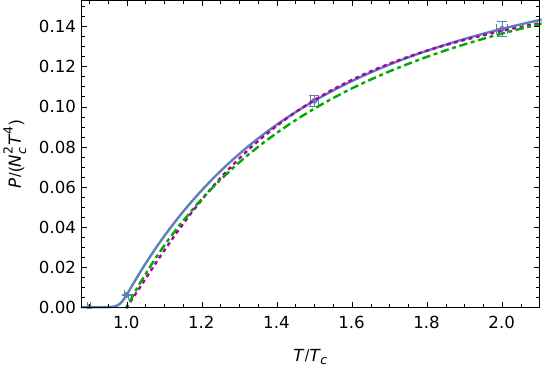}\hspace{5mm}%
\includegraphics[width=0.45\textwidth]{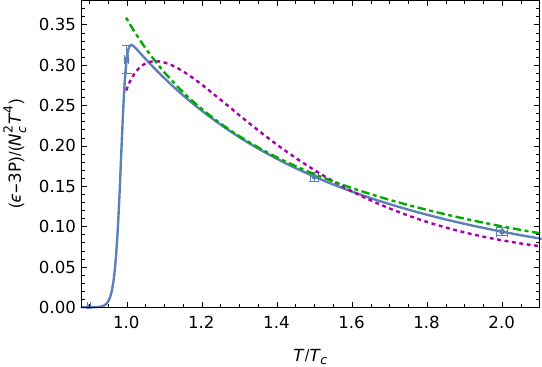}
    \caption{Comparison of the holographic thermodynamics with the potentials $V_\mathrm{II}$ and $V_\mathrm{III}$ to lattice data. The  left (right) panel shows the normalized pressure (interaction measure) as a function of the temperature. The blue curves and error bars show the lattice results for the  thermodynamics of pure Yang--Mills extrapolated to $N_c=\infty$~\cite{Panero:2009tv}. The dotted magenta and dot-dashed green curves show the holographic result using the large-$D$ thermodynamics for potentials $V_\mathrm{II}$ and $V_\mathrm{III}$ , respectively.}
    \label{fig:lattice_fit_largeD}
\end{figure}

We then discuss fitting the potentials $V_\mathrm{II}$ and $V_\mathrm{III}$. In Fig.~\ref{fig:lattice_fit_largeD} we show the results compared to large $N_c$ lattice data. Recall that for these potentials, the fitting is done using the large-$D$ thermodynamics. While the fit to the pressure (left panel) shows only a small difference between the two potentials, for the interaction measure (right panel) the differences are more pronounced. Notice that while both potentials give results rather close to the lattice data, the deviations are in opposite directions, and somewhat larger than the errors of the lattice result. The effect of having the opposite effects was our main motivation for introducing the piecewise smooth potential of Eq.~\eqref{eq:VIIIdef}, as without a kink in the potential it is difficult to create a sudden sharp transition in the functional form of the interaction measure. The deviation from lattice data is larger than the errors simply because we wanted to use relatively simple analytic formulas for the potentials, and with such simple functions it is challenging to obtain a more precise fit.

\begin{figure}\centering
\includegraphics[width=0.45\textwidth]{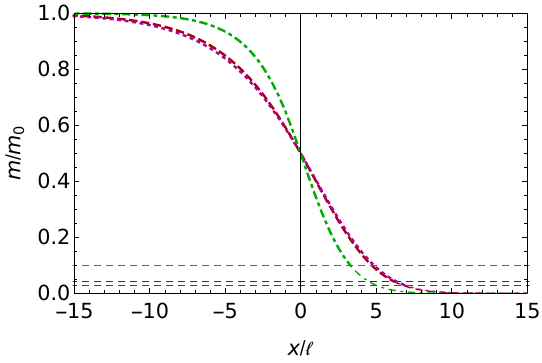}\hspace{5mm}%
\includegraphics[width=0.45\textwidth]{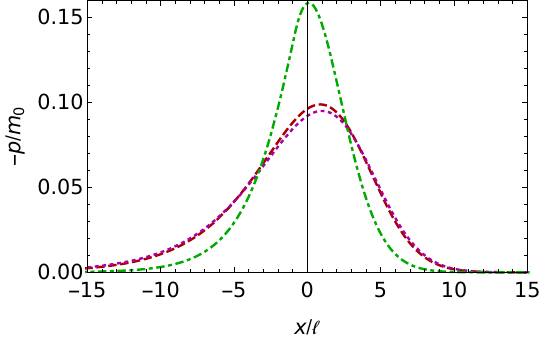}
    \caption{Profile of the domain wall solutions for different choices of potential as a function of the spatial coordinate $x$. Left: profile of the mass density $m$. Right: profile of the momentum density $p$. The dashed red, dotted magenta and dot-dashed green curves show the result using the large-$D$ thermodynamics for potentials $V_\mathrm{I}$, $V_\mathrm{II}$, and $V_\mathrm{III}$ , respectively.}
    \label{fig:domain_wall}
\end{figure}

\subsection{Properties of the domain wall}

We then construct the static domain wall solutions following the analysis of Sec.~\ref{sec:dw}. The profiles of the domain wall solutions, normalized to the value $m_0$ of $m$ in the deconfined phase, are shown in Fig.~\ref{fig:domain_wall}. The normalization is chosen because we want to show quantities that are invariant under the shift transformation~\eqref{eq:Rscale}.
We see that the results for the potential $V_\mathrm{I}$ (fitted to lattice data in the full IHQCD setup) and for the potential  $V_\mathrm{II}$ (adjusted to agree better with the data when large-$D$ thermodynamics is used) are almost identical. Given the above discussion, this suggests that the large-$D$ truncation has only a small effect on the results.
However, the potential with a kink at $\phi=0$ differs from the other potentials significantly. Again based on previous discussion, this difference signals the uncertainty in the results due to the choice of potential in the holographic approach in general, independently of the large-$D$ limit. However, since the variation of the fits in Fig.~\ref{fig:lattice_fit_largeD} is significantly larger than the lattice errors, this variation should be seen as the difference between two extreme cases rather than an estimate of a typical uncertainty due to the potential choice.

\begin{figure}\centering
\includegraphics[width=0.45\textwidth]{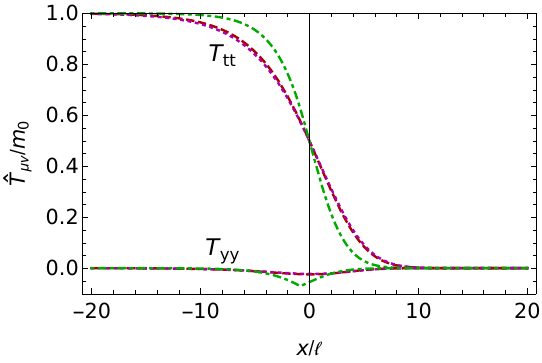}\hspace{5mm}%
\includegraphics[width=0.45\textwidth]{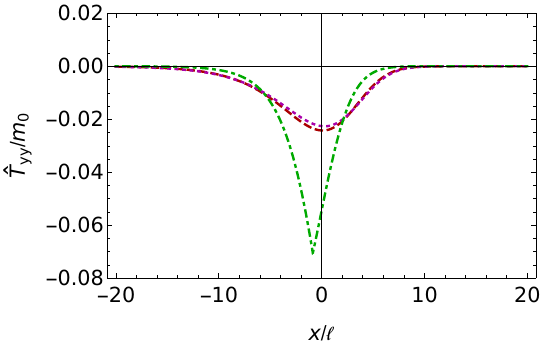}
    \caption{The components of the energy-momentum tensor over the domain wall of Fig.~\ref{fig:domain_wall}. Left: the nontrivial components $\widehat T_{\mu\nu}$ normalized to the value $m_0$ of the mass on the deconfined side. Right: Plot zoomed in to show the details of $\widehat T_{yy}$ near the center of the wall.}
    \label{fig:Tmunu}
\end{figure}

In Fig.~\ref{fig:Tmunu} we show the nontrivial components of the (normalized) energy-momentum tensor $\widehat T_{\mu\nu}$ using the expressions in~\eqref{eq:Tmunufinal}. Notice that the component perpendicular to the wall, $\widehat T_{xx}$, as well as nondiagonal components, vanish identically. It follows directly from the conservation of $\widehat  T_{\mu\nu}$ that the pressure $\widehat  T_{xx}$ must be constant,  reflecting mechanical equilibrium, and the fact that pressures of different sides must balance for the wall to be static. In our conventions, the constant is zero, because we set the pressure of the confined phase to zero in this section, corresponding to choosing the lower limit of the integration in~\eqref{eq:Idef} to be at $u=+\infty$.
The energy density $\widehat T_{tt}$, shown in the left panel, is proportional to $m(x)$ shown in Fig.~\ref{fig:domain_wall}. The component parallel to the wall, $\widehat T_{yy}$, is strongly suppressed with respect to $\widehat T_{tt}$ as shown by the left panel. The dependence on $x$ of this component is shown more clearly in the right panel. It roughly follows the behavior of $p(x)$ in Fig.~\ref{fig:domain_wall}.

\subsection{Matching with Yang--Mills theory and comparison to lattice results}

Finally, we want to make contact with the physical units given by the Yang--Mills theory that controls gluonic interactions in nature. The most straightforward way to do this is to use the temperature given in~\eqref{eq:temperature}. Since in this expression the $\deltaV$-term is treated as a small perturbation, and we are currently studying the IHQCD setup where the confinement-deconfinement transition takes one to the black hole phase near the minimum of the temperature where $\deltaV$ is indeed small, it makes sense to identify the leading term in~\eqref{eq:temperature} as the critical temperature of the Yang--Mills theory. Note that comparing our definition in~\eqref{eq:deltaVKdef} to~\eqref{eq:potential} we have effectively set $n=1$ (so that there is no explicit expansion parameter when we are applying the large-$D$ approximation). Therefore the identification boils down to
\be \label{eq:Tcritid}
 T_c \approx \frac{1}{4\pi \ell} \ .
\ee
Inserting here the critical temperature of Yang--Mills theory $T_c \approx 270$~MeV~\cite{Boyd:1996bx,Lucini:2005vg}, obtained by comparing the lattice result for the Yang--Mills string tension to the real-world value, we obtain
\be \label{eq:ellHagedorn}
 \ell \approx 0.058~\text{fm} \ ,
\ee
which gives a physical interpretation for the length scale in the plots of the domain wall in Figs.~\ref{fig:domain_wall} and~\ref{fig:Tmunu}. The scale is significantly shorter than that directly given by the intrinsic scale $\sim 200$~MeV of the theory. This is due to the factor of $4\pi$ in~\eqref{eq:Tcritid}.

Another way to extract the physical value of $\ell$ is to match the geometry between the exact numerical solution of the IHQCD model (see, e.g.,~\cite{Gursoy:2009jd}) and the expressions found by expanding in $1/n$. After this matching has been established, one can extract the physical value by using the comparison of the IHQCD results to lattice data. The fact that $\ell$ was defined through the normalization of the potential in~\eqref{eq:V0} suggests that one should compare the expressions for the potential evaluated on the solutions. As it turns out, an efficient way to do this is to use the combination
 \be
  e^{2A}V(\phi)
 \ee
which after inserting the Ansatz from Sec.~\ref{sec:model} becomes
 \be
  e^{2A}V(\phi)  \approx \frac{n^2}{\ell^2}\left[1+\frac{1}{n}\deltaV \right] \exp\left[\frac{2}{n}A_\mathrm{s}^{(1)}\right]
 \ee
with corrections suppressed in $1/n$.

\begin{figure}\centering
\includegraphics[width=0.7\textwidth]{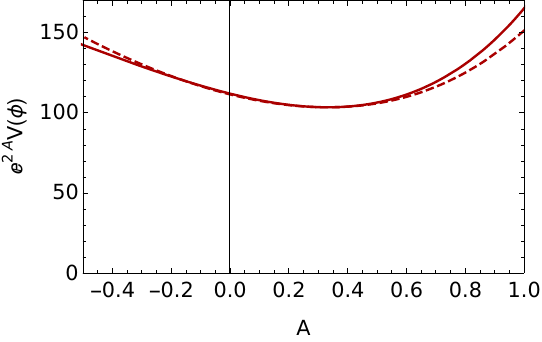}
    \caption{The comparison of the combination $e^{2A}V(\phi)$ as a function of the scale factor $A$ evaluated on the solution in IHQCD (solid red curve) and in the large-$D$ framework (dashed red curve) for the potential $V_\mathrm{I}$.  }
    \label{fig:VA_matching}
\end{figure}

We compare this combination, $e^{2A}V(\phi)$, between IHQCD and the large-$D$ framework in Fig.~\ref{fig:VA_matching}, as a function of the scale factor $A$. We used $V_\mathrm{I}$ from~\eqref{eq:VIdef}. The scale factor  in IHQCD is defined as in~\eqref{eq:gds2}, and solved numerically, whereas in the large-$D$ framework we use the expression in~\eqref{eq:AAnsatz}. On the large-$D$ side, we set $\phi_0=0$ and fitted the values of $A_0$ and $\ell$ such that good agreement near the minimum of the potential with the IHQCD result was found. Fixing the units such that the critical temperature in IHQCD is $270$~MeV, the value of $\ell$ becomes
\be \label{eq:ellgeometry}
 \ell \approx 0.065~\text{fm}\ ,
\ee
i.e., a value comparable to~\eqref{eq:ellHagedorn}.

Having obtained the physical value for $\ell$, we can extract the widths of the domain walls described by the solutions in Fig.~\ref{fig:domain_wall} in physical units. A possible definition is
\be
 \Delta = -\frac{m_0}{m'(x_{1/2})} \ ,
\ee
where $m(x_{1/2})=m_0/2$. Using the value~\eqref{eq:ellHagedorn} yields
\be
\label{eq:Deltares}
  \Delta = \left\{
  \begin{array}{ccc}
   0.61~\text{fm} &\quad & (\text{potential}\ V_\mathrm{I})  \\
   0.63~\text{fm} &\quad & (\text{potential}\ V_\mathrm{II})  \\
   0.37~\text{fm} &\quad & (\text{potential}\ V_\mathrm{III})
  \end{array}
  \right. \ .
\ee
Alternatively, we may use the definition
\be
 \Delta_{0.1} = x_{0.1}-x_{0.9} \ ,
\ee
with $m(x_{0.1})=0.1m_0$ and $m(x_{0.9})=0.9m_0$. This gives slightly higher numbers,
\be
\label{eq:Deltares2}
  \Delta_{0.1} = \left\{
  \begin{array}{ccc}
   0.65~\text{fm} &\quad & (\text{potential}\ V_\mathrm{I})  \\
   0.68~\text{fm} &\quad & (\text{potential}\ V_\mathrm{II})  \\
   0.43~\text{fm} &\quad & (\text{potential}\ V_\mathrm{III})
  \end{array}
  \right. \ .
\ee
As one also sees directly from Fig.~\ref{fig:domain_wall}, the actual width of the domain wall is relatively large as compared to the value of $\ell$: the obtained values for the width are only slightly below $1/T_c$.

We can also insert the result in Fig.~\ref{fig:Tmunu} in the formula~\eqref{eq:surfacetension} to obtain numerical estimates for the surface tension of Yang--Mills theory. However, as seen from the formula, this requires knowing the value of the Newton constant $G_5$ in our action~\eqref{eq:EDaction}. This value can be obtained from the fits to lattice data by using the large-$D$ thermodynamics, shown in Figs.~\ref{fig:lattice_fit_ihqcd} and~\ref{fig:lattice_fit_largeD}. For example, the combination for the pressure shown in these figures is given by
\be
 \frac{P}{T^4N_c^2} = \frac{\mathcal{K}}{n^4(1+\delta V/n)^4} \widehat T_{xx} = \frac{\mathcal{K}}{2n^4(1+\delta V/n)^4} I
\ee
where we used~\eqref{eq:temperature},~\eqref{eq:Tmunuphys}, and~\eqref{eq:Tmunufinal}. The  dimensionless proportionality coefficient is
\be
 \mathcal{K} = \frac{(4\pi)^4e^{3A_0}\ell^3}{16\pi G_5 N_c^2} \ .
\ee
Note that here the combination $G_5 N_c^2$ is natural because in this combination the dependence on $N_c$ cancels at large $N_c$. The large-$D$ fits to lattice data give the numbers
\be
 \mathcal{K} = 0.277 \quad (\text{potential}\ V_\mathrm{I}) \ , \quad
  \mathcal{K} = 0.230 \quad (\text{potential}\ V_\mathrm{II}) \ , \quad
    \mathcal{K} = 0.337 \quad (\text{potential}\ V_\mathrm{III}) \ .
\ee
When computing these numbers, as above, we set $n=1$ so that $\deltaV$ plays the role of the expansion parameter.

Instead of the surface tension $\sigma$, the lattice analyses typically quote the value of the dimensionless combination  $\sigma/T_c^3$ or $\sigma/(T_c^3N_c^2)$~\cite{Lucini:2005vg,Rindlisbacher:2025dqw}.
By using~\eqref{eq:surfacetension},~\eqref{eq:temperature} and~\eqref{eq:Tmunufinal} we can write the expression taking into account the large $N_c$ scaling  as
\be
 \frac{\sigma}{T_c^3N_c^2} = \frac{\mathcal{K}}{4\pi n^{7/2}(1+\delta V_c/n)^3} \int_{-\infty}^\infty d \left(\frac{x}{\ell}\right)\, \frac{p(x)^2}{m(x)} \ ,
\ee
where $\deltaV_c$ refers to the value at the critical temperature of large-$D$ thermodynamics. Note that, as it turns out, the physical value of $\ell$ is not needed to evaluate this formula, as only the dimensionless ratio $x/\ell$ appears.
Inserting the numerical solutions and value of $\mathcal{K}$ in this formula, setting again $n=1$, we find
\be \label{eq:tensionresult}
  \frac{\sigma}{T_c^3N_c^2} = \left\{
  \begin{array}{ccc}
   0.0057 &\quad & (\text{potential}\ V_\mathrm{I})  \\
   0.0050 &\quad & (\text{potential}\ V_\mathrm{II})  \\
   0.0095 &\quad & (\text{potential}\ V_\mathrm{III})
  \end{array}
  \right. \ .
\ee
These numbers should be compared to the lattice result~\cite{Rindlisbacher:2025dqw},
\be \label{eq:latticetension}
 \frac{\sigma}{T_c^3N_c^2} = 0.0182(7)-\frac{0.194(15)}{N_c^2}\qquad  (\text{lattice}) \ .
\ee
While the numbers are in the same ballpark, our holographic large-$D$ results are clearly smaller than the large $N_c$ lattice number, i.e., the leading term in the lattice fit of~\eqref{eq:latticetension}. However, the $1/N_c^2$ correction in the lattice fit brings the lattice result closer\footnote{Setting $N_c=4$ in this asymptotic formula gives $\sigma/(T_c^3 N_c^2)\approx 0.006$. However, note that setting $N_c=3$, which may be outside the region of validity of the formula, leads to a negative value.} to our results at finite $N_c$.

We remark that~\cite{Rindlisbacher:2025dqw} also gives a result for the latent heat at the transition,
\be \label{eq:latticelatent}
 \frac{L}{T_c^4N_c^2} = 0.360(6)-\frac{1.88(17)}{N_c^2}\qquad  (\text{lattice}) \ .
\ee
This number should be compared to the discontinuity of the interaction measure in our fit in Fig.~\ref{fig:lattice_fit_largeD} (right). The recent precision result~\eqref{eq:latticelatent} is rather high compared to the lattice data we are using in the fit. Correcting for this mismatch would slightly alleviate the tension between the lattice results  and our result for the surface tension.

Finally, we stress that the differences between potentials $V_\mathrm{I}$ and $V_\mathrm{II}$ defined in~\eqref{eq:VIdef} and~\eqref{eq:VIIdef}, respectively, are relatively close for all observables in~\eqref{eq:Deltares},~\eqref{eq:Deltares2} and~\eqref{eq:tensionresult}, and the same is true for the domain wall profiles in Figs.~\ref{fig:domain_wall} and~\ref{fig:Tmunu}. As we argued above, this suggests that the large-$D$ approximation is working. At the same time, the larger differences to the results obtained with potential $V_\mathrm{III}$ indicate that uncertainties in the construction of the holographic model in general (independently of the large-$D$ approach), are larger than the effect of the large-$D$ approximation.

\section{Conclusions}

In this article, we studied a framework for dynamical black hole solutions in five-dimensional Einstein-dilaton gravity established in~\cite{Jarvinen:2024vdi}. This framework arises as a systematic expansion, motivated by an analogous expansion in $1/D$ in large-dimensional Einstein gravity~\cite{Emparan:2015gva} with $D$ denoting the number of dimensions, dimensionally reduced to five dimensions. We restricted our study to the leading nontrivial order in this expansion, and the properties of the consequent four-dimensional effective description of the black holes, which generalizes the large-$D$ analysis~\cite{Emparan:2015gva,Emparan:2025yfy} in previous literature.

The purpose of our article was twofold: we both provided more details and background for the results already presented in~\cite{Jarvinen:2024vdi}, and extended this study towards new directions. To be precise, we provided additional details about the derivation of the four-dimensional effective equations describing the black hole dynamics (see Sec.~\ref{sec:BHET}), including full solutions of the geometry and its asymptotics. We also extended the discussion of the thermodynamics to include the full expressions for the energy-momentum tensor in Sec.~\ref{sec:BHET}, and extended the discussion of domain walls to spherical and cylindrical cases in Sec.~\ref{sec:dw}. Moreover, we worked out in more detail the application of our framework to the IHQCD model: apart from solving the domain wall, we analyzed the surface tension and presented an improved comparison of the model and the results to lattice results for the Yang--Mills theory in Sec.~\ref{sec:application}. As completely new directions, we discussed non-equilibrium-steady-state interfaces in Sec.~\ref{sec:dw} and completion of the model with nontrivial renormalization group flows near the boundary in Sec.~\ref{sec:UV}.

Perhaps the most pressing open question is how well our framework applies to the description of physically interesting theories such as the Yang--Mills theory. The applicability of the large-$D$-motivated expansion is restricted to a specific class of potentials defined in~\eqref{eq:potential}. This choice was argued to represent theories with nearly critical dynamics with respect to confinement, e.g., theories which show a weak finite-temperature deconfinement transition. Previous holographic analysis has shown that Yang--Mills theory is  in this class~\cite{Gursoy:2007cb,Gursoy:2007er,Gubser:2008ny}. And indeed, as demonstrated in~\cite{Jarvinen:2024vdi}, applying the large-$D$ framework to IHQCD yields promising results: The specific potential of the framework~\eqref{eq:potential} can be matched with the IHQCD potentials such that the terms suppressed or enhanced at large D are indeed numerically suppressed or enhanced. Moreover, simple consistency checks such as adjusting the potential and observing the variations in the results work out nicely. In addition, as we observed here in Sec.~\ref{sec:application}, directly comparing the large-$D$ solution to IHQCD geometry and Yang--Mills data also works, suggesting that using the leading-order large-$D$ estimates leads to errors of about ten per cent\footnote{As we argued above, the errors can be estimated via differences between potentials $V_\mathrm{I}$ and $V_\mathrm{II}$ in Figs.~\ref{fig:domain_wall} and~\ref{fig:Tmunu}, and between the results for the tension in Eq.~\eqref{eq:tensionresult}, as well as via the difference between the values of $\ell$ between different fits in Eqs.~\eqref{eq:ellHagedorn} and~\eqref{eq:ellgeometry}.} at most.

However, while these observations are encouraging, they are somewhat indirect tests: Because only homogeneous static solutions for IHQCD geometries are currently available, we cannot do a direct comparison of our results to nontrivial geometries or time-dependent solutions in IHQCD. Consequently, we are lacking precise quantitative estimates of the validity of the large-$D$ expansion. This may be improved by future work. One way to obtain precise estimates would be to construct numerically such geometries, starting from the standard static domain wall between confined and deconfined phases,  directly in IHQCD, and compare to our results. This is a somewhat challenging numerical task, but having the approximate solution obtained in our framework could make it easier to find the solution. A second way to obtain quantitative estimates about the validity of the expansion would be to compute results to higher order in $1/D$, and observe how fast they converge~\cite{Suzuki:2015axa,Emparan:2018bmi}.

Our analysis of the near-boundary completion in Sec.~\ref{sec:UV} revealed other points about the application to IHQCD that require further attention. First, the analysis of Sec.~\ref{sec:application} and that of~\cite{Jarvinen:2024vdi} showed that one can keep the $\deltaV$ term of the potential~\eqref{eq:potential} to be small, as required by the large-$D$ expansion, by restricting to relevant values of the dilaton field. However, the coefficient $V_0$ in~\eqref{eq:potential} should be large as $D$ grows, and there seems to be no mechanism to obtain a large value for this parameter. This is confirmed by the near-boundary completion procedure, which appears to produce $\mathcal{O}(1)$ values for $V_0$. Nevertheless, we found the value of $V_0$ to be numerically large when comparing to IHQCD potentials, which may be accidental. Second, the matching procedure of Sec.~\ref{sec:UV} required a parametrically long range of solution where the large-D approach works, meaning that the $\deltaV$ term in the potential remains small. Given our potentials in Fig.~\ref{fig:deltaV}, it is not obvious that this can be obtained in our IHQCD setups. Checking this requires a more extensive numerical study, where one compares numerical solutions for the fluctuations to the near-horizon black hole solutions explicitly.

There are various other future directions to explore. On the one hand, one may analyze and extend the framework discussed in this article. There is ongoing work to extend the framework to charged black holes, for which large-$D$ effective theories are known~\cite{Bhattacharyya:2015fdk,Emparan:2016sjk}. One could also try to find the solutions and effective equations to higher order in $1/D$, which, as we discussed above, could help to obtain quantitative estimates of the validity of the framework in practical applications. Another extension would be to consider space-time dependent source terms within the framework~\cite{Andrade:2018zeb,Emparan:2025yfy}. Actually, this extension could potentially help to improve the matching required for the near-boundary completion discussed in Sec.~\ref{sec:UV}. Namely, a long overlap region was required in order to suppress an undesired term in the fluctuated near-boundary solution. Such a term corresponds exactly to dynamical sources in the effective black hole dynamics of our framework. Therefore including sources could help to reduce the length of the overlap region, potentially making the matching for IHQCD easier.

On the other hand, one may work on applications of the current setup. Apart from the static or steady solutions here, it would be interesting to study dynamical, time-dependent setups, using the potentials of Sec.~\ref{sec:application} or other potentials. An interesting application would be to study turbulence through numerical simulations, which is known to simplify in the large-$D$ limit~\cite{Rozali:2017bll}. Dynamics of phase transitions could also be studied via simulations, addressing configurations such as expansions and collisions of nucleated bubbles (of a thermodynamically dominant phase within a subdominant phase) in the presence of a first-order phase transition (see~\cite{Bantilan:2020pay,Bea:2022mfb} for such simulations in full holographic models). Eventually, production of gravitational waves could be computed using results from these simulations~\cite{Bigazzi:2020avc,Ares:2021nap,Bea:2021zol}. In such studies, one can use the models for the dynamics of the Yang--Mills theory constructed here, i.e., the IHQCD setup of Sec.~\ref{sec:application}, or other setups such as models with a phase transition between two black hole phases, which may be simpler to simulate.

\section*{Acknowledgments}
We thank Tuna Demircik, Wanxiang Fan, Maurizio Firrotta, Romuald Janik, Niko Jokela, Jichao Li, Li Li, Edwan Pr\'eau, Jacob Sonnenschein, Javier Subils, and Ryotaku Suzuki for discussions.
DW was supported by an INFN postdoctoral fellowship and the INFN project ST\&FI “String Theory and Fundamental Interactions”. LBP was supported by UKRI5072 grant number 3109736.

\appendix

\section{Dimensional reduction} \label{app:dimred}

We discuss here some details of dimensional reduction from high-dimensional Einstein gravity to a dilaton gravity in a lower number of dimensions~\cite{Kanitscheider:2008kd,Gouteraux:2011qh}. Starting from the Einstein-Hilbert action with a cosmological constant in $D = d + n$ dimensions,
\begin{equation} S_{D} = \int d^d x\, d^n y\sqrt{-G}\left(R - 2\Lambda\right)\,. \end{equation}

One can consider warped product spaces having a metric of the form
\begin{equation} ds^2 = e^{2\delta_1 \phi(x)} g_{\mu\nu}(x)dx^\mu dx^\nu + e^{2\delta_2 \phi(x)}h_{ab}(y)dy^a dy^b\,, \label{eq:geometry_dimred}
\end{equation}
where $\mu$, $\nu = 0,1,\ldots,d-1$ and $a$, $b = 1,2,\ldots,n$. Note that we do not need to assume that $n$ is large at this stage.

The last equation defines also the dilaton field $\phi(x)$ in this context. The reduction becomes simple if we assume the metric $h_{ab}$ to be that of a space of constant curvature, such that the Ricci tensor and scalars associated with it are:
\begin{equation} R^{(h)}_{ab} = \frac{s (n-1)}{\ell^2} h_{ab}\,,\qquad R^{(h)} = \frac{s n(n-1)}{\ell^2}\,. \end{equation}
Here the choice $s=0$ will correspond to flat space, $s=1$ to the sphere, and $s=-1$ to hyperbolic space.
 
By straightforward computations, the Einstein equations in the full $D$-dimensional space can then be written as:
\begin{equation} R_{\mu\nu} - \frac12 g_{\mu\nu} R = \frac12 g_{\mu\nu} V(\phi)-\frac{4}{d-2}\left(\frac12 g_{\mu\nu}(\pa\phi)^2 - \pa_\mu\phi \pa_\nu \phi\right)\,, \label{eq:reducedeom1} \end{equation}
\begin{equation} \Box \phi + \frac{d-2}{8}V^\prime(\phi) = 0 \,,\label{eq:reducedeom2} \end{equation}
provided that we choose\footnote{A different choice of parameters will introduce extra terms into the equations of motion, such as $(\pa \phi)^2$ in \eqref{eq:reducedeom2}.}
\begin{equation} \delta_1 = \frac{2\sqrt{n}}{(d-2)\sqrt{n+d-2}}\,,\qquad \delta_2 = -\frac{2}{\sqrt{n(n+d-2)}}\,. \end{equation}

Note that the covariant derivatives and curvature tensor in the equations of motion are computed using only the $d$-dimensional metric $g_{\mu\nu}(x)$.

The dilaton potential obtained from the reduction, as defined by eq.~\eqref{eq:reducedeom1}, is:
\begin{equation} \label{eq:reducedfullV}
V(\phi) = -2\Lambda e^{2\beta\phi} + \frac{s n(n-1)}{\ell^2} e^{2\beta^\prime\phi}\,,
\end{equation}
with
\begin{equation} \beta = \delta_1\,,\qquad \beta^\prime =\frac{n+d-2}{n} \delta_1\,. \end{equation}

Eqs.~\eqref{eq:reducedeom1}--\eqref{eq:reducedeom2} are exactly the equations of motion that follow from the $d$-dimensional Einstein-dilaton action \eqref{eq:EDaction}, which (with  $d=5$) is the starting point of our model,
\begin{equation}
	S_\mathrm{red} = \frac{1}{16\pi G_d}\int d^d x \sqrt{-g} \left(R - \frac{4}{d-2}(\pa\phi)^2 + V(\phi)\right)
\end{equation}

The potential~\eqref{eq:reducedfullV} generally has two exponential terms, one sourced by the $D$-dimensional cosmological constant, and another by the curvature of the internal space spanned by the $y$-coordinates. Assuming now that $n$ is large and setting $d=5$, the exponents are given by the expansion
\begin{equation} 2\beta = \frac43 - \frac{2}{n} + \ldots\,,\qquad 2\beta^\prime = \frac43 + \frac{2}{n} +\ldots\,.
\end{equation}
It is important that in the $n\to\infty$ limit, both choices yield $V(\phi) = V_0 e^{\frac43\phi}$. The leading $1/n$ corrections following from this reduction are linear, in the sense that they determine the correction to the potential to be $\deltaV(\phi) \propto\phi$.

From here we can immediately recover the results of the effective theories of~\cite{Emparan:2015gva} from our general equations. In the flat space theory $\Lambda=0$ where the inner space is spherical ($s=1$), we get the potential $\deltaV(\phi) = 2\phi - 1$. Another simple possibility~\cite{Caldarelli:2013aaa} yields the AdS effective theory, when $\Lambda<0$ and the inner space is flat ($s=0$), resulting in $\deltaV(\phi) = -2\phi + 7$. Note that the constants in $\deltaV$ originate from the expansion of $V_0$ rather than the exponent.

\section{Solution of the geometry at next-to-leading order}\label{app:nlosolution}

In this Appendix, we write down explicitly the solution to the next-to-leading order corrections in $1/n$, governed by equations~\eqref{eq:A1storder}--\eqref{eq:algebraic}.

The solution of~\eqref{eq:As1storder} may be written as
\begin{align}
 A_\mathrm{s}^{(1)}({\cal R},t,x) &= c_4(t,x)-\frac{1}{4
   m(t,x)}\int_0^{\cal R}\frac{d\rs}{\rs^2}\log \left(\frac{1-\rs m(t,x)}{1-{\cal R} m(t,x)}\right) \delta V'\left(\frac{1}{2} \log \rs\right) & \nonumber \\
   &-\frac{\log (1- {\cal R} m(t,x)) \left(-c_2(t,x)+c_3(t,x)+\ell \pa_ip^i(t,x)+ m(t,x) c_1(t,x)\right)}{2
   m(t,x)} &
\end{align}
After imposing the horizon regularity condition~\eqref{eq:regcond}, the solution simplifies to
\begin{align}
 A_\mathrm{s}^{(1)}({\cal R},t,x) &= c_4(t,x)-\frac{\log (1- {\cal R} m(t,x))}{4
   m(t,x)}\int_{\cal R}^{1/m(t,x)}\frac{d\rs}{\rs^2} \delta V'\left(\frac{1}{2} \log \rs\right) & \nonumber\\
   &-\frac{1}{4
   m(t,x)}\int_0^{\cal R}\frac{d\rs}{\rs^2}\log \left(1-\rs m(t,x)\right) \delta V'\left(\frac{1}{2} \log \rs\right) &
   \label{eq:Assolreg}
\end{align}

For generic potential corrections $\deltaV$, the integrals above may have divergences at $\rs=0$. Note that in particular the large-$D$ constructions of Appendix~\ref{app:dimred}, which produce $\deltaV$'s linear in the dilaton, lead to such divergences.
We can regularize the integral by introducing an arbitrary parameter $\epsilon$, where the integral from $\rs=0$ to $\rs=\epsilon$ is separated as the divergent part with the regular part starting from $\rs=\epsilon$. We then expand the divergent part and absorb all the divergence into all the $c(t,x)$ with redefinitions. The regularized solution is
\begin{align} \label{eq:As1reg}
 A_\mathrm{s}^{(1)}({\cal R},t,x) &= c_4(t,x)-\frac{\log (1- {\cal R} m(t,x))}{4
   m(t,x)}\int_{\cal R}^{1/m(t,x)}\frac{d\rs}{\rs^2} \delta V'\left(\frac{1}{2} \log \rs\right) & \nonumber\\
   &-\frac{1}{4
   m(t,x)}\int_\epsilon^{\cal R}\frac{d\rs}{\rs^2}\log \left(1-\rs m(t,x)\right) \delta V'\left(\frac{1}{2} \log \rs\right)+\text{div1}(t,x)\,,
\end{align}
with the divergent part
\begin{equation}
    \text{div1}(t,x)=\frac{1}{4}\int_0^\epsilon \frac{d\rs}{\rs}\left(1+\frac{m(t,x)\rs}{2}+ \cdots\right)\deltaV'\left(\frac{1}{2}\log{\rs}\right)
\end{equation}
where in the expansion we should include all terms that are divergent, depending on the near-boundary asymptotics of the potential. Alternatively, we could also keep the full expression or use the expansion to all orders, in which case $\epsilon$ does not need to be small. Below we keep the terms in the integrands up to order $\morder{u^0\deltaV}$ which is enough to cancel the divergences for typical (i.e., power-like) potential choices.
The divergence can be absorbed into $c_4$ by redefining
\begin{equation}
    c_4(t,x)=c_4'(t,x)-\frac{1}{4}\int_0^\epsilon \frac{d\rs}{\rs}\left(1+\frac{m(t,x)\rs}{2}\right)\deltaV'\left(\frac{1}{2}\log{\rs}\right)\,.
\end{equation}

For the other functions, we only write the solutions where the regularity condition~\eqref{eq:regcond} has been imposed.
The solution for the next-to-leading order correction to the scale factor reads
\begin{align} \label{eq:A1reg}
 &A^{(1)}({\cal R},t,x) = \frac{1}{3} \left[c_1(t,x)-2 c_4(t,x)\right]\log {\cal R} +c_5(t,x)-\frac{p(t,x)^2}{6
   m(t,x)}{\cal R} & \nonumber\\
   &+\frac{1}{6 m(t,x)}\int _\epsilon^{\cal R}\frac{d\rs}{\rs^2}\left[\log \left(\frac{{\cal R}}{\rs}\right) \log (1-\rs m(t,x))+\text{Li}_2({\cal R} m(t,x))-\text{Li}_2(\rs m(t,x))\right]  & \nonumber\\
 &\phantom{====}\times \delta V'\left(\frac{1}{2} \log \rs\right)-\frac{\text{Li}_2({\cal R} m(t,x))}{6 m(t,x)} \int_\epsilon^{1/m(t,x)}\frac{d\rs}{\rs^2} \deltaV'\left(\frac{1}{2} \log \rs\right)  +\text{div2}(t,x) \,.
\end{align}
Note here that the divergences in most of the terms involving the polylog $\mathrm{Li}_2$ cancel. The rest of the divergences are included in
\begin{equation}
\begin{split}
       \text{div2}(t,x)=&\frac{1}{6}\int_0^\epsilon \frac{d\rs}{\rs}\left[\left(1+\frac{m(t,x)\rs}{2}\right)\log \rs-\left(1+\frac{m(t,x)\rs}{4}\right)\right.\\
       &\left.-\log {\cal R}\left(1+\frac{m(t,x)\rs}{2}\right)\right]\deltaV'\left(\frac{1}{2} \log \rs\right)\,, 
\end{split}
\end{equation}
which can be absorbed into $c_5$ by the redefinition
\begin{equation}
    c_5(t,x)=c_5'(t,x)-\frac{1}{6}\int_0^\epsilon \frac{d\rs}{\rs}\left[\left(1+\frac{m(t,x)\rs}{2}\right)\log \rs-\left(1+\frac{m\rs}{4}\right)\right]\deltaV'\left(\frac{1}{2} \log \rs\right) \,.
\end{equation}
Note that $c_1$ does not need to be redefined. This is consistent with the fact that its defining equation in~\eqref{eq:A1storder} does not involve potential-dependent integrals.

Finally, we have the (regularized) solution
\begin{align} \label{eq:f1reg}
 &f^{(1)}({\cal R},t,x) = 2 c_1(t,x)-2 c_4(t,x)+\big[-m(t,x) c_1(t,x)+2 m(t,x) c_4(t,x) &\nonumber\\
 &+3 m(t,x)
   c_5(t,x)+c_2(t,x)\big]{\cal R} + \big[m(t,x) c_1(t,x)-2 m(t,x) c_4(t,x)\big]{\cal R} \log {\cal R} &\nonumber\\
   &+( \log {\cal R}- 1) \ell \pa_ip^i(t,x) {\cal R} -\frac12 {\cal R}^2 p_i(t, x)^2 & \nonumber\\
   &+\frac{1-{\cal R}m(t,x)}{2m(t,x)}\int_\epsilon^{\cal R}\frac{d\rs}{\rs^2}\log \left(\frac{1-\rs m(t,x)}{1-{\cal R} m(t,x)}\right) \delta V'\left(\frac{1}{2} \log \rs\right) &\nonumber\\
   &+\frac{1}{2} \left(\frac{1-{\cal R}m(t,x)}{m(t,x)}\log (1-{\cal R} m(t,x))+{\cal R}-{\cal R} \text{Li}_2({\cal R} m(t,x))\right) \int_\epsilon^{1/m(t,x)}\frac{d\rs}{\rs^2} \delta V'\left(\frac{1}{2} \log \rs\right)& \nonumber\\
   &+\frac{\cal R}{2}\int _\epsilon^{\cal R}\frac{d\rs}{\rs^2}\left(\log \left(\frac{{\cal R}}{\rs}\right) \log (1-\rs m(t,x))+\text{Li}_2({\cal R} m(t,x))-\text{Li}_2(\rs m(t,x))\right)  & \nonumber\\
 &\phantom{====}\times \delta V'\left(\frac{1}{2} \log \rs\right)-{\cal R} \int_\epsilon^{\cal R} \frac{d\rs}{\rs^2}\delta V\left(\frac{1}{2} \log \rs\right)  +\text{div3}(t,x),
\end{align}
with the divergent part being
\begin{align}
    \text{div3}(t,x)=&\int_0^\epsilon d\rs\left(-\left(\frac{m(t,x)}{4}+\frac{1}{2\rs}\right)\deltaV'\left(\frac{1}{2}\log \rs\right)\right.\nonumber\\
    &+{\cal R}\left[-\frac{1}{\rs^2}\deltaV\left(\frac{1}{2}\log \rs\right)\right.\nonumber\\
    &\left.+\left(\frac{m(t,x)^2}{8}+\frac{1}{2\rs^2}+\frac{m(t,x)}{2\rs}\left(1+\frac{m(t,x)\rs}{2}\right)\log \rs\right)\deltaV'\left(\frac{1}{2}\log \rs\right)\right]\nonumber\\
    &\left.- {\cal R} \log {\cal R}\left(\frac{m(t,x)}{2\rs}\left(1+\frac{m(t,x)\rs}{2}\right)\right)\deltaV'\left(\frac{1}{2}\log \rs\right)\right).
\end{align}
This divergence can be consistently absorbed into $c_2$ with redefinition
\begin{align}
    c_2(t,x)=&c_2'(t,x)-2m(t,x)c_4(t,x)-3m(t,x)c_5(t,x)\nonumber\\
    &-\int_0^\epsilon d\rs\left[-\frac{1}{\rs^2}\deltaV\left(\frac{1}{2}\log \rs\right)\right.\nonumber\\
    &\left.+\left(\frac{m(t,x)^2}{8}+\frac{1}{2\rs^2}+\frac{m(t,x)}{2\rs}\left(1+\frac{m(t,x)\rs}{2}\right)\log \rs\right)\deltaV'\left(\frac{1}{2}\log \rs\right)\right],
\end{align}
where we also use the redefinitions of $c_4$ and $c_5$ from the previous solutions.

\section{Near-boundary asymptotics of the solution}\label{app:UVasymptotics}

In this Appendix we write down the near-boundary asymptotics of the geometry and the dilaton field by using the solutions from Appendix~\ref{app:nlosolution}. We give first the general treatment and then discuss the special case of large-$D$ AdS geometry dimensionally reduced to five dimensions.

Consider first the case where $\delta V$ vanishes fast in the UV. Then the source terms of $A_\mathrm{s}^{(1)}$,  $A^{(1)}$, and $f^{(1)}$ are those involving $c_1$, $c_4$, and $c_5$, as can be immediately read off from Eqs.~\eqref{eq:As1reg},~\eqref{eq:A1reg}, and~\eqref{eq:f1reg}. Therefore we will set them to zero, 
\be
 c_1 = c_4 = c_5 = 0 \ .
\ee
The VEV terms read
\begin{align}
    A_\mathrm{s}^{(1)}({\cal R},t,x) &= \frac{1}{4}I(t,x){\cal R} + \mathcal{O}\left({\cal R}^2\right)  & \\    
    A^{(1)}({\cal R},t,x) &= 
    \left[-\frac{1}{6}I(t,x)-\frac{p_i(t,x)^2}{6m(t,x)}\right]{\cal R} + \mathcal{O}\left({\cal R}^2\right) & \\
    f^{(1)}({\cal R},t,x) &= \ell \pa_ip^i(t,x){\cal R}\log{\cal R} +\left[c_2(t,x)-\ell \pa_ip^i(t,x)\right]{\cal R}+ \mathcal{O}\left({\cal R}^2\right)
\end{align}
where %
\be \label{eq:Idefapp}
 I(t,x) \equiv \int_0^{1/m(t,x)}\frac{d\rs}{\rs^2} \delta V'\left(\frac{1}{2} \log \rs\right)
\ee
and $c_2$ is recognized as the next-to-leading order variation of $m$.
These expansions imply that
\be
 \phi^{(1)}({\cal R},t,x) = \frac{3}{2}\left[A_\mathrm{s}^{(1)}({\cal R},t,x)-A^{(1)}({\cal R},t,x)\right]= \left[\frac{5}{8}I(t,x)+\frac{p_i(t,x)^2}{4m(t,x)}\right]{\cal R} + \mathcal{O}\left({\cal R}^2\right)
\ee

Let us then analyze the general case, where the terms from the potential affect the asymptotic formulas. Using the exact expressions from Appendix~\ref{app:nlosolution}, the general asymptotic formulas can be found to be 
\begin{align} \label{eq:Asbdrygen}
    A_\mathrm{s}^{(1)}({\cal R},t,x) &=\frac{1}{4}\int^\calR \frac{d\rs}{\rs}\deltaV'\left(\frac{1}{2}\log\rs\right)\left[1+\morder{\calR}\right]+c_4' & \nonumber\\
    &+ \frac{\calR}{4}\int_\calR^\Lambda \frac{d\rs}{\rs^2}\deltaV'\left(\frac{1}{2}\log\rs\right)\left[1+\morder{\calR}\right] + \frac{1}{4}I_\Lambda(t,x){\cal R} + \mathcal{O}\left({\cal R}^2\right)  & \\  
    \label{eq:Abdrygen}
    A^{(1)}({\cal R},t,x) &= -\frac{1}{6}\int^{\cal R}\frac{d\rs}{\rs}\left[\log \left(\frac{{\cal R}}{\rs}\right) +1\right]\deltaV'\left(\frac{1}{2}\log\rs\right)\left[1+\morder{\calR}\right]\nonumber & \\
    &+\frac{1}{3}(c_1-2c_4')\log \calR +c_5'- \frac{\calR}{6}\int_\calR^\Lambda \frac{d\rs}{\rs^2}\deltaV'\left(\frac{1}{2}\log\rs\right)\left[1+\morder{\calR}\right] & \nonumber\\
    &+\left[-\frac{1}{6}I_\Lambda(t,x)-\frac{p_i(t,x)^2}{6m(t,x)}\right]{\cal R} + \mathcal{O}\left({\cal R}^2\right) & \\
    f^{(1)}({\cal R},t,x) &= \left[\deltaV\left(\frac 12 \log \calR\right) -\frac{1}{2}\int^\calR \frac{d\rs}{\rs}\deltaV'\left(\frac{1}{2}\log\rs\right)\right]\left[1+\morder{\calR}\right] +2c_1-2c_4' &\nonumber \\
    &+ \left[\ell \pa_ip^i(t,x)+ m(t,x)(c_1-2c_4')\right]{\cal R}\log{\cal R} & \nonumber\\
    &+\left[c_2'(t,x)-m(t,x)c_1-\ell \pa_ip^i(t,x)\right]{\cal R}+ \mathcal{O}\left({\cal R}^2\right) \,,  \label{eq:fbdrygen}
\end{align}
where
\be \label{eq:ILambdadef}
 I_\Lambda(t,x) \equiv \int_\Lambda^{1/m(t,x)}\frac{d\rs}{\rs^2} \delta V'\left(\frac{1}{2} \log \rs\right)
\ee
with $\Lambda$ being a positive cutoff.

Here we wrote separately the potential contributions to the source and VEV terms. In the source term integrals (first lines in each expression) we did not specify the lower limit of the integral since their variations can be absorbed in redefinitions of the integration constants $c_4'$ and $c_5'$, which no longer can be set to zero. However, we required them to be independent of $t$ and $x^k$ because they give contributions to the source terms. %

Note that the potential-dependent source terms are present even in the vacuum solution where $m(t,x)=0=p_i(t,x)$. That is, even the vacuum solution contains nonzero contributions in the $1/n$ expansion, except when $\deltaV = 0$. 

In the VEV terms of $A_\mathrm{s}^{(1)}$ and $A^{(1)}$ the potential UV divergence of the integral $I(t,x)$ was regulated by isolating it in a ``trivial'' piece which is independent of $t$ and $x^k$. This was done by introducing a cutoff $\Lambda$. The dependence on the cutoff cancels in the full expression.

\subsection{Expansions for dimensionally reduced AdS} \label{app:dim_red_AdS}

Since the general expressions for the solutions and for the near-boundary expansions are rather cumbersome, it is instructive to consider a simple example. A relevant example containing a UV divergence is provided by the dimensionally reduced AdS$_D$ geometry~\cite{Gouteraux:2011qh,Jarvinen:2024vdi}.

Starting from a $n+5$ -dimensional Einstein gravity with a cosmological constant $V_0$, the potential in~\eqref{eq:potential} is given by
\be
 V(\phi) = V_0e^{\frac{4\sqrt{n}}{3\sqrt{n+3}}\phi} = e^{4\phi/3}\frac{n^2}{\ell^2}\left[1+\frac{1}{n}(-2\phi+7)+\morder{\frac{1}{n^2}}\right] \ .
\ee
Here we defined the radius $\ell$ through $V_0 = (n+4)(n+3)/\ell^2$, which reduces to the expression in the main text at large $n$.
That is, we have that $\deltaV(\phi) = -2\phi+7$ and $\phi_0=0$.  Note that this is precisely the choice where our setup boils down to the earlier large-$D$ analysis of black holes in Einstein gravity~\cite{Emparan:2015gva}. Usually the constant factor in $\deltaV$ is omitted because it does not enter the effective equations. Here we keep it for full consistency. You can also use the symmetry~\eqref{eq:Rscale} to eliminate it, but this adds an overall factor to the metric.

The vacuum solution for this potential is the dimensional reduction of the higher-dimensional AdS space, and given for any value of $n$ by
\be \label{eq:AdSmetric}
 ds^2 = \left(\frac{r+\ell}{\ell}\right)^{-\frac{2}{3}(n+3)}\left(-dt^2-2dtdr+\frac{1}{n}\delta_{ij}dx^idx^j\right)
\ee
with the dilaton solution
\be
 \phi(r) = \frac{1}{2}\sqrt{n(n+3)}\,\log \frac{r+\ell}{\ell}\,.
\ee
Here we used translations in $r$ to place the boundary of the solution at $r=-\ell$. This shift is required for the solution to have smooth behavior in the limit $n\to \infty$. Taking this limit at fixed $\calR = e^{nr/\ell}$, we find that
\begin{align} \label{eq:AlargeD}
 A(r) &=  -\frac{1}{3}\log \calR +\frac{1}{6n}\left[\left(\log\calR\right)^2-6\log \calR\right] + \morder{\frac{1}{n^2}} & \\
  \phi(r) &= \frac{1}{2}\log \calR +\frac{1}{4n}\left[-\left(\log\calR\right)^2+3\log \calR\right] + \morder{\frac{1}{n^2}}&\\ A(r)&+\frac{2}{3}\phi(r) = -\frac{1}{2n}\log \calR + \morder{\frac{1}{n^2}} \ . & \label{eq:AslargeD}
\end{align}

These results can now be compared to the general asymptotics in Eqs.~\eqref{eq:Asbdrygen}--\eqref{eq:fbdrygen}. Inserting $\deltaV(\phi)=-2\phi+7$, and marking the unspecified lower endpoint of the integrals by $\delta$, we find that the source terms in the scale factors $A_\mathrm{s}^{(1)}$ and  $A^{(1)}$ become
\begin{align} \label{eq:Assrc}
A_\mathrm{s}^{(1)}\Big|_\mathrm{source}& = \frac{1}{4}\int_\delta^\calR \frac{d\rs}{\rs}\deltaV'\left(\frac{1}{2}\log\rs\right)+c_4' = -\frac{1}{2}\log\frac{\calR}{\delta}+c_4' \\    
A^{(1)}\Big|_\mathrm{source}&=-\frac{1}{6}\int_\delta^{\cal R}\frac{d\rs}{\rs}\left[\log \left(\frac{{\cal R}}{\rs}\right) +1\right]\deltaV'\left(\frac{1}{2}\log\rs\right)+\frac{1}{3}(c_1-2c_4')\log \calR +c_5' & \nonumber\\
 &= \frac{1}{6}\left[\left(\log\frac{\calR}{\delta}\right)^2+2\log\frac{\calR}{\delta}\right]+\frac{1}{3}(c_1-2c_4')\log \calR+c_5' 
\end{align}
whereas the integral in $f^{(1)}$ involves a cancellation:
\be \label{eq:fsrc}
f^{(1)}\Big|_\mathrm{source}  = \deltaV\left(\frac 12 \log \calR\right) -\frac{1}{2}\int_\delta^\calR \frac{d\rs}{\rs}\deltaV'\left(\frac{1}{2}\log\rs\right)+2c_1 -2c_4' = 7-\log\delta+2c_1 -2 c_4' \ .
\ee

We should then compare the subleading terms in the expressions in~\eqref{eq:AlargeD} and in~\eqref{eq:AslargeD} to the sources~\eqref{eq:Assrc}--\eqref{eq:fsrc}.
However, the source expressions are still not complete. Namely, the UV-divergent integrals with $\Lambda$-cutoff in~\eqref{eq:Asbdrygen} and in~\eqref{eq:Abdrygen} also contribute to the source terms. Since we are checking the vacuum solution, we set $m(t,x) = 0$ and we may also take for simplicity $\Lambda \to \infty$ which removes the VEV $I_\Lambda$. The remaining integral becomes
\be
 \calR \int_\calR^\infty \frac{d\rs}{\rs^2}\deltaV'\left(\frac{1}{2}\log\rs\right) = -2 \,,
\ee
contributing a constant factor $-1/2$ ($+1/3$) to the source term of $A_\mathrm{s}^{(1)}$ ($A^{(1)}$).
After adding these contributions, and setting $c_1=-3$, $c_4'=-(\log \delta)/2+1/2$ and $c_5' = -(\log \delta)^2/6+(\log \delta)/3-1/3$ the sources match with those of the dimensionally reduced AdS geometry, see~\eqref{eq:AlargeD} and~\eqref{eq:AslargeD}. In particular, the source for $f^{(1)}$ vanishes as it should because the empty AdS metric contains no blackening factor. %

To conclude this Appendix, let us make a simple observation about the VEV terms. For static black holes, the VEV terms in the expansions~\eqref{eq:Asbdrygen} and~\eqref{eq:Abdrygen} only involve the integral~\eqref{eq:ILambdadef} which was recognized as the pressure in~\eqref{eq:pressure}. Inserting in this definition the AdS potential $\deltaV(\phi) = -2 \phi + 7$, and taking again $\Lambda \to \infty$, which appears to be the only natural choice for this potential, we obtain
\be
 T_{xx} = \frac{e^{3A_0}}{16\pi G_5 \ell}  m \,.
\ee
Therefore, comparing to the energy density from Sec.~\ref{sec:BHET}, we have that $T_{xx} = T_{tt}/n$, which is the expected equation of state for large-$D$ black holes.

\section{Fefferman--Graham coordinates and the energy-momentum tensor} \label{app:FG}

In this Appendix, we derive the expressions for the energy-momentum tensor from the near-boundary asymptotics of the geometry by using the holographic dictionary. As it turns out, it is simpler to do this by using the analogue of the $D$-dimensional geometry, following the picture of dimensional reduction discussed in Appendix~\ref{app:dimred}. That is, as we discussed in the main text, even though we are working with five-dimensional geometries in this article, our setup is motivated by dimensional reduction of $D$-dimensional gravity in the limit of large $D$. This motivation allows us to define an ``uplifted'' $D$-dimensional geometry~\cite{Kanitscheider:2009as} through Eq.~\eqref{eq:geometry_dimred}, even if our generalized setup (where the potential variation $\delta V$ of~\eqref{eq:potential} is arbitrary) cannot be obtained from $D$-dimensional gravity via consistent dimensional reduction.

It is known that in pure Einstein-gravity the holographic dictionary implies that the energy-momentum tensor can be computed by evaluating the Brown-York stress tensor near the boundary of the geometry~\cite{Brown:1992br,Balasubramanian:1999re}. This is the standard approach at large $D$, see, e.g.,~\cite{Andrade:2020ilm}. However, in our case applying this formalism with the uplifted geometry from~\eqref{eq:geometry_dimred} is somewhat cumbersome, because it requires dealing with the full $D$-dimensional geometry, including the $D-5$ dimensional space of transverse directions that are integrated out in the reduction. Nevertheless, it is clear that if this is done, the analysis of~\cite{Andrade:2020ilm} applies directly to our case: we can always choose a $\delta V$ that behaves near the boundary such that the same asymptotics as in this reference is produced, and choosing a more general $\delta V$ only adds divergent source terms (see Appendix~\ref{app:UVasymptotics}) the effect of which will be removed by holographic renormalization.

We expect that one can also compute the energy momentum tensor directly in five-dimensions by using holographic renormalization of the Einstein-dilaton gravity, which has been worked out in detail in~\cite{Papadimitriou:2011qb}. However, the general procedure is somewhat involved, and the fact that the five-dimensional geometry is not asymptotically AdS makes the renormalization procedure in our case particularly difficult~\cite{Kanitscheider:2008kd}. We do not attempt to carry out such analysis here.

Therefore, in this Appendix, we present a somewhat more mundane approach, which only uses five-dimensional geometries: we transform the five-dimensional part of the uplifted geometry to Fefferman--Graham coordinates, which allows us to read the energy-momentum tensor directly from the asymptotics of the geometry~\cite{deHaro:2000vlm,Skenderis:2002wp}. Interestingly, we shall see that this approach leads to a minor mismatch with respect to the results obtained in \cite{Andrade:2020ilm}, see the comment on this in Sec.~\ref{sec:Tmunu}.

We only consider the case where $\deltaV$ vanishes so fast near the boundary that the $\deltaV$-dependent source terms (see Appendix~\ref{app:UVasymptotics}) can be set to zero. Note that $\delta_1$ in~\eqref{eq:geometry_dimred} equals $2/3$ in the limit of large $n$. Using this definition, our Ansatz for the metric from Sec.~\ref{sec:model}, and the near-boundary expansions from Appendix~\ref{app:UVasymptotics}, we see that the solution for the ``uplifted'' geometry in the Eddington--Finkelstein coordinates may be written as
\begin{align}
&\left(g_{\mu\nu}^{(\phi)}\right) \equiv e^{\frac{4}{3}\phi({\cal R},t,x)}\left(g_{\mu\nu}\right) \\
&  \nonumber
\approx e^{\frac{4}{3}\phi_0+2A_0}\left(
 \begin{array}{ccccc}
    0  & -\frac{\ell}{n \cal R}  &0  &0  &0    \\
    -\frac{\ell}{n \cal R}  & -1  &0  &0  &0    \\
    0  & 0  &\frac1n  &0  &0    \\
    0  & 0  &0  &\frac1n  &0    \\
    0  & 0  &0  &0  &\frac1n     
 \end{array} \right) &\\
&  +
  e^{\frac{4}{3}\phi_0+2A_0}{\cal R}\left(
 \begin{array}{ccccc}
    0  & -\frac{\ell}{n^2 \cal R}v_{tr}(t,x)  &0  &0  &0    \\
    -\frac{\ell}{n^2 \cal R}v_{tr}(t,x)  & -V_{tt}({\cal R},t,x)
    &-\frac1n p_1(t,x)&-\frac1n p_2(t,x)  &-\frac1n p_3(t,x)    \\
    0  & -\frac1n p_1(t,x)  &\frac1{n^2}v_{11}(t,x)  &\frac1{n^2}v_{12}(t,x)  &\frac1{n^2}v_{13}(t,x)    \\
    0  & -\frac1n p_2(t,x)  &\frac1{n^2}v_{12}(t,x)  &\frac1{n^2}v_{22}(t,x) & \frac1{n^2}v_{23}(t,x)   \\
    0  & -\frac1n p_3(t,x)  &\frac1{n^2}v_{13}(t,x)  &\frac1{n^2}v_{23}(t,x)  &\frac1{n^2}v_{33}(t,x)     
 \end{array} \right) &
\end{align}
where the order of coordinates is $({\cal R},t,x^1,x^2,x^3)$, corrections are suppressed by $1/n$ and $\cal R$, and we introduced shorthand $v_{ij}$ and $V_{tt}$ for the VEV terms. We kept the leading nontrivial order of each nonzero VEV term, except for the $tt$ VEV term, which is the only $\morder{1}$ VEV. It contains a logarithmic term at subleading order,
\be
 V_{tt}({\cal R},t,x) =v_{tt}(t,x)+ \frac1n v_{tt}^{(l)}(t,x)\log{\cal R}+\frac1n v_{tt}^{(1)}(t,x) \,.
\ee
Interestingly, the five-dimensional part of the uplifted metric $g_{\mu\nu}^{(\phi)}$ also equals the string frame metric in the five-dimensional Einstein-dilaton gravity picture.

Comparing with Appendix~\ref{app:UVasymptotics}, the expressions for the VEVs are given by
\begin{align}
    v_{tr} &= \frac{1}{2} I  \,, \qquad v_{tt} = -m \,, \qquad v_{tt}^{(l)} = \ell\, \pa_i p^i \,,\qquad v_{tt}^{(1)} = \frac12 I + c_2 -\ell\, \pa_ip^i & \\
    v_{ij} & = \frac{1}{2} I \, \delta_{ij} + \frac{p_ip_j}{m} \,,
\end{align}
where $I(t,x)$ is defined in~\eqref{eq:Idefapp} in  Appendix~\ref{app:UVasymptotics}.

The geometry needs to be transformed to the Fefferman--Graham form in order to read off the energy-momentum tensor. To do this, we first transform from Eddington--Finkelstein time coordinate to ordinary time, defined through $t = \hat t -\frac{\ell}{n}\log{\cal R}$. After this we have
\begin{align} \label{eq:geomexpthat}
    &e^{-\frac{4}{3}\phi_0-2A_0}\left(g_{\mu\nu}^{(\phi)}\right) \approx \left(
 \begin{array}{ccccc}
    \frac{\ell^2}{n^2 {\cal R}^2}  & 0  &0  &0  &0    \\
    0  & -1  &0  &0  &0    \\
    0  & 0  &\frac1n  &0  &0    \\
    0  & 0  &0  &\frac1n  &0    \\
    0  & 0  &0  &0  &\frac1n     
 \end{array} \right) &\\
 &+{\cal R}\left(
 \begin{array}{ccccc}
    \frac{\ell^2}{n^2 {\cal R}^2}m(\hat t,x)  & -\frac{\ell}{n{\cal R}}m(\hat t,x)  &\frac{\ell}{n^2{\cal R}}p_1(\hat t,x)  &\frac{\ell}{n^2{\cal R}}p_2(\hat t,x)  &\frac{\ell}{n^2{\cal R}}p_3(\hat t,x)    \\
   -\frac{\ell}{n{\cal R}}m(\hat t,x) & -V_{tt}({\cal R},\hat t,x)-\frac{\ell}{n}\pa_t m(\hat t,x)\log{\cal R}
    &-\frac1n p_1(\hat t,x)&-\frac1n p_2(\hat t,x)  &-\frac1n p_3(\hat t,x)    \\
    \frac{\ell}{n^2{\cal R}}p_1(\hat t,x)  & -\frac1n p_1(\hat t,x)  &\frac1{n^2}v_{11}(\hat t,x)  &\frac1{n^2}v_{12}(\hat t,x)  &\frac1{n^2}v_{13}(\hat t,x)    \\
    \frac{\ell}{n^2{\cal R}}p_2(\hat t,x)  & -\frac1n p_2(\hat t,x)  &\frac1{n^2}v_{12}(\hat t,x)  &\frac1{n^2}v_{22}(\hat t,x) & \frac1{n^2}v_{23}(\hat t,x)   \\
    \frac{\ell}{n^2{\cal R}}p_3(\hat t,x)  & -\frac1n p_3(\hat t,x)  &\frac1{n^2}v_{13}(\hat t,x)  &\frac1{n^2}v_{23}(\hat t,x)  &\frac1{n^2}v_{33}(\hat t,x)     
 \end{array} \right) &\nonumber
\end{align}
where we inserted $v_{tt}=-m$. Note that in this expression the power counting is clear: The diagonal source term for the $\calR$ coordinate on the first line of~\eqref{eq:geomexpthat} is suppressed by $1/n^2$ due to the definition of $\calR$ in~\eqref{eq:calRdef} and the diagonal spatial components are suppressed by $1/n$ due to rescaling $z^k = x^k/\sqrt{n}$ of the spatial coordinates with respect to the physical coordinates $z^k$. In the VEV terms on the second line one encounters the same suppression factors, but there is an additional factor of $1/\sqrt{n}$ ($1/n$) in each component of the metric with one spatial index (two spatial indices). These latter factors indicate that even the physical values of the VEVs are suppressed in the limit of large $n$.

In~\eqref{eq:geomexpthat} the source term is readily in the desired FG form if we use $r$ instead of $\calR$. You might expect that it should be multiplied by the AdS factor $\sim \ell^2/r^2$ in this form. However, this factor is actually subleading. The correct way to add it in the large-$D$ limit is after a shift (see Appendix~\ref{app:UVasymptotics}) so that it becomes
\be
 \frac{\ell^2}{(r+\ell)^2} = 1 - \frac{2}{n}\log \calR+\morder{\frac{1}{n^2}}
\ee
which means that the AdS factor only appears as corrections suppressed by $1/n$. %
However, the VEVs are not yet in the desired form because of the nonzero components $g_{\cal RR}$, $g_{t{\cal R}}$, and $g_{i{\cal R}}$. To eliminate them, it turns out that we need to redefine all coordinates near the boundary. We write
\be
  {\cal R} = \widetilde{\cal R} + \epsilon_{\cal R}(\tilde t,\tilde x)\widetilde{\cal R}^2 \,,\qquad \hat t  = \tilde t + \frac{\ell}{n}\epsilon_{t}(\tilde t,\tilde x)\widetilde{\cal R} \,,\qquad x^i = \tilde x^i + \frac{\ell}{n}\epsilon_{x}^i(\tilde t,\tilde x)\widetilde{\cal R}
\ee
Inserting this to the metric, and requiring the vanishing of all VEV components in the $\cal R$ column and row at leading nontrivial order in $1/n$ gives
\be
 \epsilon_{\cal R}(\tilde t,\tilde x) = -\frac{1}{2}m(\tilde t,\tilde x) \,,\qquad \epsilon_{t}(\tilde t,\tilde x) = -m(\tilde t,\tilde x) \,, \qquad \epsilon_{x}^i(\tilde t,\tilde x) = -p^i(\tilde t,\tilde x) + \frac{\ell}{2} \pa^i m(\tilde t,\tilde x) \ .
\ee
In terms of the redefined coordinates, the metric becomes %
\begin{align}
\nonumber
    &e^{-\frac{4}{3}\phi_0-2A_0}\left(g_{\mu\nu}^{(\phi)}\right) \approx \left(
 \begin{array}{ccccc}
    \frac{\ell^2}{n^2 \widetilde \calR^2}  & 0  &0  &0  &0    \\
    0  & -1  &0  &0  &0    \\
    0  & 0  &\frac1n  &0  &0    \\
    0  & 0  &0  &\frac1n  &0    \\
    0  & 0  &0  &0  &\frac1n     
 \end{array} \right) &\\
 &+\widetilde \calR\left(
 \begin{array}{ccccc}
    0  & 0  &0  &0  &0    \\
   0 & \widehat T_{tt}(\tilde t,\tilde x)+\frac{1}{n}\widehat T_{tt}^{(1)}(\widetilde \calR,\tilde t,\tilde x)
    &\frac1n \widehat T_{t1}(\tilde t,\tilde x)&\frac1n \widehat T_{t2}(\tilde t,\tilde x)  &\frac1n  \widehat T_{t3}(\tilde t,\tilde x)    \\
    0  & \frac1n  \widehat T_{t1}(\tilde t,\tilde x)  &\frac1{n^2}\widehat T_{11}(\tilde t, \tilde x)  &\frac1{n^2}\widehat T_{12}(\tilde t, \tilde x)  &\frac1{n^2}\widehat T_{13}(\tilde t, \tilde x)    \\
    0  & \frac1n  \widehat T_{t2}(\tilde t,\tilde x)  &\frac1{n^2}\widehat T_{12}(\tilde t, \tilde x)  &\frac1{n^2}\widehat T_{22}(\tilde t, \tilde x) & \frac1{n^2}\widehat T_{23}(\tilde t, \tilde x)   \\
    0  & \frac1n  \widehat T_{t3}(\tilde t,\tilde x)  &\frac1{n^2}\widehat T_{13}(\tilde t, \tilde x)  &\frac1{n^2}\widehat T_{23}(\tilde t, \tilde x)  &\frac1{n^2}\widehat T_{33}(\tilde t, \tilde x)
 \end{array} \right) &
\end{align}
where the subleading correction to the $tt$ component is 
\begin{align}
 \widehat T_{tt}^{(1)}(\widetilde \calR,\tilde t,\tilde x) &= -v_{tt}^{(1)}(\tilde t,\tilde x)-v_{tt}^{(l)}(\tilde t,\tilde x) \log\widetilde \calR-2 \ell \pa_t v_{tt}(\tilde t,\tilde x)+ \ell \pa_t v_{tt}(\tilde t,\tilde x) \log\widetilde \calR & \\
 &= -\frac12 I(\tilde t,\tilde x) - c_2(\tilde t,\tilde x)+\ell\pa_ip^i(\tilde t,\tilde x)+2 \ell \pa_t m(\tilde t,\tilde x) & \\
 &
 -\ell \pa_ip^i(\tilde t,\tilde x) \log\widetilde \calR- \ell \pa_t m(\tilde t,\tilde x) \log\widetilde \calR \,. &
 \label{eq:logRterm}
\end{align}
Actually the tildes in the VEV term could be dropped because the difference in the  original and transformed coordinates is subleading in $\calR$. Note that using the energy conservation equation~\eqref{eq:effeq1} the coefficient of the logarithmic term in~\eqref{eq:logRterm} may be further simplified to
\be
 -\ell \pa_ip^i(\tilde t,\tilde x) - \ell \pa_t m(\tilde t,\tilde x) = -\ell^2 \pa_i\pa^i m(\tilde t,\tilde x)\,.
\ee

The leading components (in $1/n$) of the energy momentum tensor are
\begin{align} \label{eq:Tmunufinalapp}
  \widehat T_{tt}(\tilde t,\tilde x) &= -v_{tt}(\tilde t,\tilde x) = m(\tilde t,\tilde x) \,,\qquad \widehat T_{ti}(\tilde t,\tilde x) = -p_i(\tilde t,\tilde x)+\ell \pa_i m(\tilde t,\tilde x) &\\
 \widehat T_{ij}(\tilde t,\tilde x) & = v_{ij}(\tilde t,\tilde x) -\ell \pa_i p_j(\tilde t,\tilde x)-\ell \pa_j p_i(\tilde t,\tilde x) + \ell^2 \pa_i\pa_j m(\tilde t,\tilde x) & \nonumber \\
&=\frac{1}{2} I(\tilde t,\tilde x) \, \delta_{ij} + \frac{p_i(\tilde t,\tilde x)p_j(\tilde t,\tilde x)}{m(\tilde t,\tilde x)}-\ell \pa_i p_j(\tilde t,\tilde x)-\ell \pa_j p_i(\tilde t,\tilde x) + \ell^2 \pa_i\pa_j m(\tilde t,\tilde x) &\nonumber
\end{align}
Note that in the latter expressions here tildes could be dropped since the tilded coordinates differ from the original ones by corrections suppressed in $1/n$ or $\cal R$. This then gives the expressions given in the main text.

\bibliographystyle{JHEP}
\bibliography{References}

\end{document}